\documentclass[aps,pre,superscriptaddress,twocolumn,10pt]{revtex4-2}

\usepackage{amsmath}
\usepackage{amssymb}
\usepackage{amsthm}
\usepackage{bbm}
\usepackage{color}
\usepackage{graphicx}[floatfix]
\usepackage{hyperref}
\usepackage[utf8]{inputenc}
\usepackage[T1]{fontenc}
\usepackage{physics}
\usepackage{times}
\usepackage{ulem}

\hypersetup{
        colorlinks=true,linkcolor=blue,citecolor=blue,
        filecolor=blue,urlcolor=blue,breaklinks=true
}

\RequirePackage{xcolor}

\newcommand{\rev}[1]{{\color{red}{#1}}}

\newcommand{\rem}[1]{{\color{gray}{#1}}}

\newcommand{\orcid}[1]{\href{https://orcid.org/#1}{\includegraphics[width=7pt]{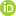}}}

\theoremstyle{definition}

\begin{document}

\title{Unitary evolution for a two-level quantum system in fractional-time scenario}

        \author{D. Cius\orcid{0000-0002-4177-1237}}
        \email{danilocius@gmail.com}
        \affiliation{
                Programa de P\'os-Gradua\c{c}\~{a}o em Ci\^{e}ncias/F\'{i}sica,
                Universidade Estadual de Ponta Grossa,
                84030-900 Ponta Grossa, Paran\'a, Brazil
        }

        \author{L. Menon Jr.\orcid{0000-0001-7057-8789}}
        \affiliation{
                Departamento de F\'{i}sica,
                Pontif\'{i}cia Universidade Cat\'{o}lica do Rio de Janeiro,
                22451-900 Rio de Janeiro, Rio de Janeiro, Brazil
        }

        \author{M. A. F. dos Santos\orcid{0000-0001-8735-0617}}
        \affiliation{
                Departamento de F\'{i}sica,
                Pontif\'{i}cia Universidade Cat\'{o}lica do Rio de Janeiro,
                22451-900 Rio de Janeiro, Rio de Janeiro, Brazil
        }

        \author{{A. S. M. de Castro\orcid{0000-0002-1521-9342}}}
        \affiliation{
                Programa de P\'os-Gradua\c{c}\~{a}o em Ci\^{e}ncias/F\'{i}sica,
                Universidade Estadual de Ponta Grossa,
                84030-900 Ponta Grossa, Paran\'a, Brazil
        }
        \affiliation{
                Departamento de F\'{i}sica,
                Universidade Estadual de Ponta Grossa,
                84030-900 Ponta Grossa, Paran\'{a}, Brazil
        }

        \author{Fabiano M. Andrade\orcid{0000-0001-5383-6168}}
        \affiliation{
                Programa de P\'os-Gradua\c{c}\~{a}o em Ci\^{e}ncias/F\'{i}sica,
                Universidade Estadual de Ponta Grossa,
                84030-900 Ponta Grossa, Paran\'a, Brazil
        }
        \affiliation{
                Departamento de Matem\'{a}tica e Estat\'{i}stica,
                Universidade Estadual de Ponta Grossa,
                84030-900 Ponta Grossa, Paran\'{a}, Brazil
        }

        \date{\today}

\begin{abstract}
The time-evolution operator obtained from the fractional-time Schr\"{o}dinger equation (FTSE) is said to be non-unitary since it does not preserve the norm of the vector state in time. As done in the time-dependent non-Hermitian quantum formalism, for a traceless non-Hermitian two-level quantum system, we demonstrate that it is possible to map the non-unitary time-evolution operator in a unitary one. It is done by considering a dynamical Hilbert space with a time-dependent metric operator, constructed from a Hermitian time-dependent Dyson map, in respect to which the system evolves in a unitary way, and the standard quantum mechanics interpretation can be made properly. To elucidate our approach, we consider three examples of  Hamiltonian operators and their corresponding unitary dynamics obtained from the solutions of FTSE, and the respective Dyson maps. 
\end{abstract}


\maketitle

\section{Introduction}
\label{sec:intro}

Nowadays, quantum physics in fractional calculus and its general properties have been investigated and applied in different areas such as optics \cite{longhi2015fractional,huang2017beam},  $\mathcal{PT}$-symmetric systems \cite{zhang2016pt}, and nonlinear variable-order time fractional Schrödinger equation \cite{heydari2019cardinal}. Experimental evidences of fractional quantum mechanics are found by Wu and co-workers in the investigation of spontaneous emission from a two-level atom in anisotropic one-band photonic crystals \cite{wu2010}. They demonstrated that fractional calculus circumvents the inconsistency of the unphysical bound state for the resonant atomic frequency lying outside the photonic band gap corresponding to the prolonged lifetime effect in the experimental observation, in which it would disappear when the emission peak lay
outside the band gap \cite{fujita2005}. The key ingredient in these approaches is the use of a differential operator with fractional order written as $\mathcal{D}_{z}^{\alpha}f(z)$, with $\alpha \in (0,1]$  representing the fractional order of the correspondent differential operator $\mathcal{D}_{z}$ on the variable $z$. Different definitions are considered for the fractional differential operator depending on the applied mathematical functions \cite{podlubny1998fractional}. For instance, fractional calculus in statistical physics is usually applied in continuous-time random walks to describe
transport phenomena. In this case, the diffusion equation with fractional derivative in spatial coordinate implies walkers that can perform Lévy flights.
In contrast, the diffusion with fractional-time derivative implies sub-diffusive behavior \cite{metzler2000random}. In the quantum framework, the fractional Schrödinger equation was introduced by Nick Laskin \cite{laskin:00a,laskin:00b,wei:16,laskin:16}. Instead of the second-order spatial derivative in the well-known Schrödinger equation, Laskin has considered a fractional Laplacian operator defined by a Reisz derivative \cite{laskin:00a,laskin:00b,wei:16,laskin:16}. Furthermore, Mark Naber \cite{naber:04} has proposed a 
FTSE assuming the Caputo fractional derivative in the place of the ordinary time derivative in such a way that the equation is written as
\begin{equation}
\label{FTSE}
    i^{\alpha} \hbar_{\alpha} \;^\text{C} _0\mathcal{D}_t^{\alpha}\vert\Psi^{\alpha}(t)\rangle
    =
    \hat{\mathcal{H}}_{0}^{\alpha}\vert\Psi^{\alpha}(t)\rangle \text{,}
\end{equation}
with
\begin{equation} 
\label{FD}
{}^\text{C}_{0}\mathcal{D}_t^{\alpha}f(t) =  \frac{1}{\Gamma(1-\alpha)}\int_{0}^{t}d\tau(t-\tau)^{-\alpha}\frac{d }{d\tau}f(\tau) ,
\end{equation}
defining the fractional Caputo derivative for  $\alpha\in [0,1)$. Here we may consider all the variables and Planck constant in Eq. (\ref{FTSE}) as dimensionless quantities. In Ref. \cite{naber:04}, it is argued that the imaginary unit is raised to  the same power as the time coordinate by performing a Wick
rotation. More details about this issued is discussed in Refs. \cite{naber:04,iomin2009fractional}. Furthermore, according to Ref. \cite{nasrolahpour2011electron},
there is an alternative form to write the FTSE  which consists in the exchange  $i^{\alpha} \to i$ in Eq. (\ref{FTSE}). 
However, the form in Eq. (\ref{FTSE}) for the FTSE has been more investigated in the literature.
Solutions for FTSE has been investigated in many situations, including, for instance, in the fractional dynamics of free particles
\cite{naber:04} and in the case under the influence of $\delta$ potentials \cite{lenzi2013time}. 
An interesting noteworthy is a mathematical correspondence between the FTSE and  the fractional-time diffusion equation \rev{\cite{naber:04,iomin2009fractional,nasrolahpour2011electron,lenzi2013time}}, viewed as describing a
non-Markovian process. This correspondence can be verified by replacing the real-time for the imaginary-time ($t \to i t$) in the fractional diffusion equation \cite{naber:04}. Further, the  fractional-time diffusion equation presents two different behaviors: sub- and superdiffusive behavior for the respectively fractional parameter $0<\alpha<1$ and $1<\alpha<2$ \cite{metzler2000random}. Also, a relation between classical geometric diffusion and quantum dynamics can be seen in Ref. \cite{iomin2020}. 


The Hamiltonian operator appearing in fractional-time quantum mechanics
is called pseudo-Hamiltonian operator because, in fractional-time
quantum dynamics, many undesired results appear, such as the
non-existence of stationary energy levels, non-unitarity of the
evolution operator, non-conservation of probability as we can see in
\cite{laskin2017time}.
A remarkable feature in fractional-time quantum dynamics is the
non-unitary time-evolution of the state vector of the system
$\vert\Psi^{\alpha}(t)\rangle$, a fact which reveals that the system
evolves employing a time-dependent non-Hermitian Hamiltonian operator.
This is a relevant result for our purposes and arises by applying the Riemann-Liouville derivative operator ${}^{\text{RL}} _0\mathcal{D}_t^{1-\alpha}$ on both sides of the
Eq. \eqref{FTSE}. Evoking the property of the fractional  differentiation
for $0<\alpha<1$, we have \cite{iomin2019app}
\begin{equation}
{}^{\text{RL}}_0\mathcal{D}_t^{1-\alpha}\;^\text{C} _0\mathcal{D}_t^{\alpha}   \vert\Psi^{\alpha}(t)\rangle
= \partial_{t}\vert\Psi^{\alpha}(t)\rangle,
\end{equation}
with
\begin{equation}
{}^{\text{RL}}_0\mathcal{D}_t^{\alpha} f(t) = \frac{1}{\Gamma(1-\alpha)}\frac{d}{dt} \int_{0}^{t}d\tau(t-\tau)^{-\alpha}\frac{d}{d\tau}f(\tau).
\end{equation}
As a direct consequence, Eq. \eqref{FTSE} can be rewritten in terms of a
first-order time-derivative, and the time-dependent Hamiltonian operator
appears in the second term on the right hand side of Eq. \eqref{FTSE},
leading to
\begin{align}
\label{TDSE0}
i^{\alpha}\hbar_{\alpha}\partial_{t}\vert\Psi^{\alpha}(t)\rangle
=
\hat{\mathcal{H}}_{0}^{\alpha} \;^{\text{RL}} _0\mathcal{D}_t^{1-\alpha}\vert\Psi^{\alpha}(t)\rangle .
\end{align}
It represents a non-local time in the Hamiltonian operator due to the
integral in the definition of fractional derivative [see
Eq. \eqref{FD}], implying in solutions that are not invariant under
time-reversal transformation.
There are different proposals to map the non-unitary fractional
evolution operator into a unitary one
\cite{zhang2021quantization,laskin2017time,iomin2019fractional}.
As consequences of the non-unitary are the non-conservation of
probability and the failure of the quantum superposition law.
In this work, we focus our efforts on mapping the non-unitary fractional
time-evolution operator in a unitary one by employing non-Hermitian
quantum mechanics procedures for time-dependent metrics.
This procedure allows for a proper quantum mechanical interpretation of
the fractional description concerning the modified inner product from
the non-Hermitian approach.

Non-Hermitian Hamiltonians have attracted much attention since the
seminal paper of Bender and Boettcher \cite{bender:98}.
They demonstrated that the hermiticity condition may be replaced by the
weaker requirement of time-spatial invariance ($\mathcal{PT}$-symmetry)
that guarantees the reality of spectrum, \textit{i.e.},
$[\hat{H},\mathcal{PT}]=0$. Indeed, an entirely real spectrum is ensured
when, in addition to being invariant, the Hamiltonian shares its
eigenstates with the $\mathcal{PT}$ operator, and in this case, the
system is said to have unbroken $\mathcal{PT}$-symmetry
\cite{bender:07}.
Although this conjecture has been proved to work in some cases
\cite{dorey:01}, the probabilistic interpretation assigned to quantum
states had been put into an obscure scenario.

The probabilistic interpretation of the correspondent quantum theory was
established by A. Mostafazadeh
\cite{mostafazadeh:02a,mostafazadeh:02b,mostafazadeh:02c} in the context
of pseudo-hermiticity.
The non-unitary time evolution generated by a non-Hermitian Hamiltonian
operator was circumvented by considering a Hilbert space based on a
time-independent metric operator where the non-Hermitian Hamiltonian
enjoys self-adjointness.
The so-called metric operator is Hamiltonian-dependent, and when its
spectrum is strictly positive, the non-Hermitian Hamiltonian is called
quasi-Hermitian.
In this case, a similarity transformation maps the non-Hermitian
Hamiltonian into a Hermitian one. The quasi-Hermiticity property was
investigated in earlier work \cite{scholtz:92}, where a consistent
quantum mechanical framework had been established.

Looking at time-dependent non-Hermitian quantum systems, many proposals
have been presented for reestablishing the self-adjointness
\cite{znojil:08,gong:13,fring:16a,luiz:20}.
The interpretation of this quantum theory is designed in a dynamical
Hilbert space which is constructed using a time-dependent metric
operator compared to the time-independent case.
However, in this case, the Hermitian Hamiltonian is not connected to the
non-Hermitian one through a similarity transformation but rather employs
a generalized relation called the time-dependent quasi-hermiticity
relation \cite{fring:16a}.
Considering a time-dependent metric operator implies that the
time-dependent Hamiltonian operator is not a quantum observable, and it
must be tread as a mere generator of the unitary time-evolution
\cite{fring:17}.
Other contributions have been presented regarding time-dependent
non-Hermitian Hamiltonians in
Refs. \cite{maamache:17,fring:19,khantoul:17,mana:20,koussa:20,cius:22}.

In the present work, our proposal consists in describing the non-unitary
fractional-time Schr\"{o}dinger evolution of two-level quantum systems
in the light of non-Hermitian quantum mechanics formalism taking into
account a time-dependent metric \cite{fring:16a,fring:17}.
This time-dependent metric allows us to build a dynamical Hilbert space
where the states evolve unitarily.
In this regard, this work begins with a brief review of time-dependent
non-Hermitian quantum systems in Sec. \ref{sec:TDNHH}.
In Sec. \ref{sec:Unit}, we obtain the non-unitary time-evolution
operator by solving the FTSE and we propose a time-dependent Dyson map
(related to the Hilbert space metric) that allows us to obtain an
equivalent unitary time-evolution operator which describes the dynamics
through the usual quantum mechanical formalism.
In Sec. \ref{sec:Appl} we present three simple applications to elucidate
our approach by considering Hermitian and non-Hermitian Hamiltonian
operators in the FTSE.
Our conclusions follow in Sec. \ref{sec:Conc}.


\section{Review on Time-Dependent non-Hermitian systems}
\label{sec:TDNHH}

In this section, we briefly revisit the protocol developed in
Refs. \cite{fring:16a,fring:17} to treat non-Hermitian quantum systems
governed by time-dependent Hamiltonian operators.
For this purpose, the starting point is to consider the two
time-dependent Schrödinger equations (TDSE),
\begin{subequations}
\label{TDSEs}
    \begin{align}
    \label{TDSEs-NH}
        i\hbar\partial_{t}\vert \Psi(t) \rangle &= \hat{H}(t)\vert \Psi(t) \rangle \text{,}
        \\
    \label{TDSEs-H}
        i\hbar\partial_{t}\vert \psi(t) \rangle &= \hat{h}(t)\vert \psi(t) \rangle \text{,}
    \end{align}
\end{subequations}
where the time-dependent Hamiltonian operator $\hat{H}(t)$ is not
Hermitian ($\hat{H}(t) \neq \hat{H}^\dagger(t)$) whereas the
$\hat{h}(t)$ is Hermitian ($\hat{h}(t) = \hat{h}^{\dagger}(t)$).
The states $\vert \Psi(t) \rangle$ and $\vert \psi(t) \rangle$ are
supposed to be related by means of the time-dependent invertible Dyson
map  $\hat{\eta}(t)$ according to
\begin{eqnarray}
\label{States}
    \vert \psi(t) \rangle = \hat{\eta}(t)\vert \Psi(t) \rangle.
\end{eqnarray}
From the substitution of Eq. \eqref{States} into Eq. \eqref{TDSEs}, the
two Hamiltonians become related to each other as follows
\begin{equation}
\label{hHer}
    \hat{h}(t)
    =
    \hat{\eta}(t)\hat{H}(t)\hat{\eta}^{-1}(t)
    +
    i\hbar\partial_{t}\hat{\eta}(t)\hat{\eta}^{-1}(t).
\end{equation}
As a consequence of the hermiticity of $\hat{h}(t)$, the following
equation arises
\begin{equation}
    \hat{H}^{\dagger}(t)\hat{\Theta}(t)-\hat{\Theta}(t)\hat{H}(t)
    =
    i\hbar\partial_{t}\hat{\Theta}(t),
\end{equation}
which is known as the time-dependent quasi-hermiticity relation
\cite{fring:16a}, where the metric operator $\hat{\Theta}(t)$ is
identified as
\begin{equation}
    \label{ss}
    \hat{\Theta}(t) = \hat{\eta}^\dagger(t)\hat{\eta}(t).
\end{equation}

Having evidenced the relationship between the states in \eqref{States}
for the different Hamiltonian operators, it is essential to verify how
the dynamics corresponding to each of the equations \eqref{TDSEs-NH} and
\eqref{TDSEs-H} are in correspondence.
The formal solution of the non-Hermitian Schrödinger equation
\eqref{TDSEs-NH} can be written as
\begin{equation}
\label{solU-TD}
\vert \Psi(t) \rangle = \hat{U}(t,t_0)\vert \Psi(t_0) \rangle,
\end{equation}
where $\hat{U}(t,t_0)$ is the non-unitary time-evolution operator, which
sounds inadequate to treat the quantum dynamics as established by the
principles of quantum mechanics.
On the other hand, the formal solution of the Hermitian Schrödinger
equation \eqref{TDSEs-H} reads
\begin{equation}
\label{solu-TD}
\vert \psi(t) \rangle = \hat{u}(t,t_0)\vert \psi(t_0) \rangle,
\end{equation}
with $\hat{u}(t,t_0)$ being the unitary time-evolution operator.
From the Eqs. \eqref{States}, \eqref{solU-TD} and \eqref{solu-TD}, it is
straightforward to obtain
\begin{align}
\vert \Psi(t) \rangle
&= \hat{\eta}^{-1}(t) \hat{u}(t,t_0)\vert \psi(t_0) \rangle,
\nonumber\\
&= \hat{U}(t,t_0)\hat{\eta}^{-1}(t_0)\vert \psi(t_0) \rangle.
\end{align}
This result leads us to the formal relationship between both operator
$\hat{u}$ and $\hat{U}$ written as
\begin{equation}
\label{relUu}
\hat{u}(t,t_0) = \hat{\eta}(t)\hat{U}(t,t_0)\hat{\eta}^{-1}(t_0).
\end{equation}
Therefore, the unitary time-evolution operator can be obtained from the
non-unitary one by applying the time-dependent Dyson map, which accounts
for the relationship of the dynamics associated with \eqref{TDSEs-NH}
and \eqref{TDSEs-H} in the time interval from $t_0$ to $t$.

Equations \eqref{TDSEs-NH} and \eqref{TDSEs-H} may be treated
equivalently by considering the Hermitian system characterized by a
time-dependent Hamiltonian $\hat{h}(t)$ and the usual metric operator
$\hat{I}$ (the identity operator).
In contrast, the non-Hermitian system is characterized by the
time-dependent Hamiltonian $\hat{H}(t)$ and the time-dependent metric
operator $\hat{\Theta}(t)$.
The relation between the probability densities in both Hermitian and
non-Hermitian systems is
\begin{eqnarray}
\label{IP}
    \langle \psi(t)\vert \hat{I} \vert \psi(t) \rangle = \langle \Psi(t)\vert \hat{\Theta}(t) \vert \Psi(t) \rangle,
\end{eqnarray}
naturally leading to the definition of a modified inner product denoted
by $\langle \cdot  \vert  \cdot \rangle_{\Theta(t)}$, given by
\begin{align}
  \label{IP1}
  \langle \Psi(t)  \vert  \tilde{\Psi}(t) \rangle_{\Theta(t)}  = {}
  &
    \langle \Psi(t) \vert \hat{\Theta}(t) \vert \tilde{\Psi}(t) \rangle \nonumber \\
  = {}
  & \langle \psi(t)  \vert  \tilde{\psi}(t) \rangle,
\end{align}
with $\vert \psi(t) \rangle = \hat{\eta}(t)\vert \Psi(t)\rangle$ and $\vert \tilde{\psi}(t) \rangle = \hat{\eta}(t)\vert \tilde{\Psi}(t)\rangle$ in according to Eq. \eqref{States}. 
This implies that an observable $\hat{\mathcal{O}}(t)$ in the
non-Hermitian system is related to an observable $\hat{o}(t)$ in the
Hermitian system by means of the similarity transformation
\begin{equation}
\label{Oo}
    \hat{\mathcal{O}}(t) = \hat{\eta}^{-1}(t) \hat{o}(t) \hat{\eta}(t).
\end{equation}
Consequently, the mean values of any observable become related as
\begin{align}
  \langle\hat{\mathcal{O}}(t)\rangle_{\Theta(t),\Psi(t)}
  = {}
  &
    \langle \Psi(t) \vert  \hat{\Theta}(t) \hat{\mathcal{O}}(t) \vert \Psi(t) \rangle
    \nonumber \\
     = {} &
    \langle \psi(t) \vert  \hat{o}(t) \vert \psi(t) \rangle
     \nonumber \\
    = {} &
    \langle \hat{o}(t) \rangle_{\psi(t)}.
\end{align}

A direct consequence of handling with time-dependent metric is the
non-observability of the Hamiltonian\rem{,} since $\hat{h}(t)$ and
$\hat{H}(t)$ do not satisfy Eq. \eqref{Oo} due to the additional term
$i\hbar\partial_{t}\hat{\eta}(t)\hat{\eta}^{-1}(t)$ as expressed in
Eq. \eqref{hHer}.
Indeed, the Hamiltonian $\hat{H}(t)$ does not stand for a physical
observable; therefore, there is no reason that the eigenvalues are real
at any instant.
It implies that the Hamiltonian plays the role of the time-evolution
generator and cannot be understood as the energy operator.
However, an observable corresponding to the energy operator may be
defined as \cite{fring:17}
\begin{align}
\label{eOp}
    \tilde{H}(t)
  \equiv {}
  &
    \hat{\eta}^{-1}(t) \hat{h}(t) \hat{\eta}(t)
    \nonumber \\
  = {}
  &
    \hat{H}(t) + i\hbar\hat{\eta}^{-1}(t)\partial_{t}\hat{\eta}(t),
\end{align}
which makes a clear distinction between the Hamiltonian operator
$\hat{H}(t)$ and energy operator $\tilde{H}(t)$. Notice that the energy
operator does not represent a Hamiltonian operator since it does not
satisfy both Eqs. \eqref{TDSEs-NH} and \eqref{TDSEs-H} previously
considered.

After this short review in which the time-dependent non-Hermitian
approach was briefly discussed, we present how such a formalism can be
applied directly to fractional dynamics, considering some examples of
physical interest.

\section{Unitary dynamics for fractional-time two-level quantum system}
\label{sec:Unit}

\subsection{Fractional Dynamics of Two-Level Quantum System}
\label{subsec:twolev}

In this section, we consider a non-Hermitian two-level quantum system
described by the traceless Hamiltonian operator in the form
\begin{equation}
\label{H0}
    \hat{\mathcal{H}}_{0}^{\alpha}= \hbar_{\alpha}\hat{\boldsymbol{\sigma}} \cdot \boldsymbol{\omega}^{\alpha},
\end{equation}
where
$\boldsymbol{\omega}^{\alpha}=
(\omega_{1}^{\alpha},\omega_{2}^{\alpha},\omega_{3}^{\alpha})$
is a complex three-dimensional vector field with the component
$\omega_{k}^{\alpha}=\omega_{k,\text{R}}^{\alpha}+i\omega_{k,\text{I}}^{\alpha}$,
and the indexes $\text{R}$ and $\text{I}$ denote the real and imaginary
parts, respectively.
Furthermore, the vector $\hat{\boldsymbol{\sigma}} =
(\hat{\sigma}_{1},\hat{\sigma}_{2},\hat{\sigma}_{3})$ is written in
terms of the Pauli matrices $\hat{\sigma}_{k}$ ($k=1,2,3$), which
satisfy the $SU(2)$ Lie algebra,
\begin{equation}
    [\hat{\sigma}_{3},\hat{\sigma}_{\pm}]
    =
    \pm 2\hat{\sigma}_{\pm},
    \qquad
    [\hat{\sigma}_{+},\hat{\sigma}_{-}]
    =
    \hat{\sigma}_{3},
\end{equation}
where $\hat{\sigma}_{\pm} = (\hat{\sigma}_{1} \pm i \hat{\sigma}_{2})/2
$ are the ladder operators.

In the fractional-time scenario, the dynamics of the quantum vector
state is described by the formal solution
\begin{equation}
\label{FTSEsol}
\vert \Psi^{\alpha}(t)\rangle =
\hat{U}_{\alpha}(t) \vert\Psi^{\alpha}(0)\rangle ,
\end{equation}
in which the system evolves from an initial state
$\vert\Psi^{\alpha}(0)\rangle$ to the state
$\vert \Psi^{\alpha}(t)\rangle$ through the non-unitary time-evolution
operator written as
\begin{eqnarray}
   \hat{U}_{\alpha}(t) = E_{\alpha}\left(\frac{\hat{\mathcal{H}_{0}^{\alpha}}}{i^{\alpha}\hbar_{\alpha}} \,t^{\alpha}\right),
   \qquad \hat{U}_{\alpha}^{-1}(t)\neq \hat{U}_{\alpha}^{\dagger}(t),
\end{eqnarray}
with $E_{\alpha}(x)= \sum_{k=0}^{\infty} x^{k}/\Gamma(\alpha k +1)$ as
the one-parameter Mittag-Leffler function \cite{podlubny1998fractional}.
The properties of Pauli's matrices allow rewriting $\hat{U}_{\alpha}(t)$
in the form
\begin{equation}
\label{Vmat}
\hat{U}_{\alpha}(t)
\doteq
\left[
\begin{matrix}
W_{+}^{\alpha}(t) & T_{-}^{\alpha}(t) \\
T_{+}^{\alpha}(t) & W_{-}^{\alpha}(t)
\end{matrix}
\right],
\end{equation}
where the coefficients satisfy the initial conditions
\begin{equation}
W_{\pm}^{\alpha}(0) = 1, \quad T_{\pm}^{\alpha}(0) = 0,
\end{equation}
in such a way that they can be expressed by
\begin{subequations}
        \label{WT}
        \begin{align}
        W_{\pm}^{\alpha}(t) &=
        \mathcal{C}_{\alpha}\left(\Delta_{\alpha}t^{\alpha}\right)
        \pm
        i^{-\alpha} \frac{\omega_{3}^{\alpha}}{\Delta_{\alpha}}
        \mathcal{S}_{\alpha}\left(\Delta_{\alpha}t^{\alpha}\right),
        \\
        T_{\pm}^{\alpha}(t) &=
        i^{-\alpha}\frac{\omega_{1}^{\alpha} \pm i\omega_{2}^{\alpha}}{\Delta_{\alpha}}
        \mathcal{S}_{\alpha}\left(\Delta_{\alpha}t^{\alpha}\right).
        \end{align}
\end{subequations}
Here, the complex functions
$\mathcal{C}_{\alpha}(\Delta_{\alpha}t^{\alpha})$ and
$\mathcal{S}_{\alpha}(\Delta_{\alpha}t^{\alpha})$ are given in the form
\begin{subequations}
\begin{align}
    \mathcal{C}_{\alpha}(\Delta_{\alpha}t^{\alpha})
  = {}
  &
    \sum_{k=0}^{\infty}(-1)^{\alpha k} \frac{(\Delta_{\alpha}t^{\alpha})^{2k}}{\Gamma(2k\alpha+1)}
    \nonumber\\
  = {}
  &
    \frac{E_{\alpha}(i^{-\alpha}\Delta_{\alpha}t^{\alpha}) + E_{\alpha}(-i^{-\alpha}\Delta_{\alpha}t^{\alpha})}{2},
\\
    \quad
    \mathcal{S}_{\alpha}(\Delta_{\alpha}t^{\alpha})
  = {}
  &
    \sum_{k=0}^{\infty}(-1)^{\alpha k} \frac{(\Delta_{\alpha}t^{\alpha})^{2k+1}}{\Gamma[(2k+1)\alpha+1]}
    \nonumber\\
  = {}
  &
    \frac{E_{\alpha}(i^{-\alpha}\Delta_{\alpha}t^{\alpha}) - E_{\alpha}(-i^{-\alpha}\Delta_{\alpha}t^{\alpha})}{2i^{-\alpha}},
\end{align}
\end{subequations}
which depend on the complex constant $\Delta_{\alpha} $ defined as
\begin{equation}
  \Delta_{\alpha} =
  \sqrt{(\omega_{1}^{\alpha})^{2} + (\omega_{2}^{\alpha})^{2}
    + (\omega_{3}^{\alpha})^{2}}.
\end{equation}

\subsection{Unitary time-evolution}
As done in the time-dependent pseudo-Hermitian description previously
discussed, we claim here that exists a time-dependent metric such that
\begin{equation}
\langle \Psi^{\alpha}(t)\vert \Psi^{\alpha}(t)\rangle_{\Theta_{\alpha} (t)}
=
\langle \Psi^{\alpha}(0)\vert \Psi^{\alpha}(0)\rangle_{\Theta_{\alpha} (0)},
\end{equation}
in which  $\hat{\Theta}_{\alpha}(t)=
\hat{\eta}_{\alpha}^{\dagger}(t)\hat{\eta}_{\alpha}(t)$ is the metric
operator.
To determine the parameters of the Dyson map $\hat{\eta}_{\alpha}(t)$,
we suppose that the state $\vert \Psi^{\alpha}(t)\rangle$ evolves
non-unitarily in relation to the trivial metric, and is related to a
state $\vert \psi^{\alpha}(t)\rangle$ through the equality
\begin{equation}
\label{psiPsiTDfrac}
\vert \psi^{\alpha} (t) \rangle =
\hat{\eta}_{\alpha} (t) \vert \Psi^{\alpha}(t) \rangle.
\end{equation}
This relation is in agreement with Eq. \eqref{States}, and the state is
assumed to evolve unitarily in time, driven by a unitary evolution
operator $\hat{u}_{\alpha}(t)$:
\begin{equation}
  \label{UFS1}
  \vert \psi^{\alpha} (t) \rangle =
  \hat{u}_{\alpha}(t)\vert \psi^{\alpha} (0) \rangle.
\end{equation}
In what follows, Eq. \eqref{relUu} allows us to write the unitary
time-evolution operator $\hat{u}_{\alpha}(t)$ in terms of the Dyson map
$\hat{\eta}_{\alpha}(t)$  and the non-unitary time-evolution operator
$\hat{U}_{\alpha}(t)$:
\begin{equation}
\label{uU}
\hat{u}_{\alpha}(t) = \hat{\eta}_{\alpha}(t)\hat{U}_{\alpha}(t)\hat{\eta}_{\alpha}^{-1}(0).
\end{equation}
Since the non-unitary time-evolution operator $\hat{U}_{\alpha}(t)$ is
known, we have to specify the time-dependent Dyson map parameters for
mapping the fractional dynamics in a unitary one.

The choice of time-dependent Dyson map $\hat{\eta}_{\alpha}(t)$ is not
unique, and for this reason, we propose a general Hermitian form given
by
\begin{equation}
  \label{DysonGD}
  \hat{\eta}_{\alpha}(t) =
  e^{\kappa_{\alpha}(t)}
  e^{\lambda_{\alpha}(t)\hat{\sigma}_{+}}
  e^{\ln{\Lambda_{\alpha}(t)}\hat{\sigma}_{3}/2}
  e^{\lambda_{\alpha}^{\ast}(t)\hat{\sigma}_{-}}.
\end{equation}
The parameters appearing in Eq. \eqref{DysonGD},
$\lambda_{\alpha}(t) \in \mathbb{C}$ and $\kappa_{\alpha}(t),
\Lambda_{\alpha}(t) \in \mathbb{R}$, are assumed to be time-dependent
under the additional condition $\Lambda_{\alpha}(t)>0$.
Moreover, we can also write the Dyson map as a second-order matrix in
the form
\begin{equation}
\label{etamat}
\hat{\eta}_{\alpha}(t) =        \frac{e^{\kappa_{\alpha}(t)}}{\sqrt{\Lambda_{\alpha}(t)}}\left[
        \begin{matrix}
        \Lambda_{\alpha}(t) + |\lambda_{\alpha}(t)|^{2} & \lambda_{\alpha}(t) \\
        \lambda_{\alpha}^{\ast}(t) & 1
\end{matrix} \right],
\end{equation}
which is most adequate from now on.
In applying the result of Eq. \eqref{etamat} into Eq. \eqref{uU}, and
taking into account the matrix forms of $\hat{\eta}_{\alpha}$ and
$\hat{U}_{\alpha}$, respectively prescribed by Eqs. \eqref{Vmat} and
\eqref{etamat}, we obtain the matrix form of $\hat{u}_{\alpha}$ as
\begin{equation}
\label{u_matrix}
\hat{u}_{\alpha}(t)
=
\left[
\begin{matrix}
\varpi_{+}^{\alpha}(t) & \varpi_{-}^{\alpha}(t) \\
\tau_{+}^{\alpha}(t) & \tau_{-}^{\alpha}(t)
\end{matrix}
\right].
\end{equation}
In Eq.  \eqref{u_matrix}, the coefficients $\varpi_{\pm}^{\alpha}$ and  $\tau_{\pm}^{\alpha}$ are given by
\begin{subequations}
  \label{coeu}
  \begin{align}
    \varpi_{\pm}^{\alpha}(t)  = {}
    &    \pm\frac{e^{\kappa_{\alpha}(t)-\kappa_{\alpha}(0)}}
      {\sqrt{\Lambda_{\alpha}(t)\Lambda_{\alpha}(0)}}
        \nonumber\\
    &
      \times
        \left\{
        \lambda_{\alpha}(t)\zeta_{\pm}^{\alpha}(t)
        +
        \left[\Lambda_{\alpha}(t) + |\lambda_{\alpha}(t)|^{2} \right]\xi_{\pm}^{\alpha}(t)
        \right\},
        \\
    \tau_{\pm}^{\alpha}(t)  = {}
    &
      \pm\frac{e^{\kappa_{\alpha}(t)-\kappa_{\alpha}(0)}}
      {\sqrt{\Lambda_{\alpha}(t)\Lambda_{\alpha}(0)}}
        \left[
        \lambda_{\alpha}^{\ast}(t) \xi_{\pm}^{\alpha}(t)
        + \zeta_{\pm}^{\alpha}(t)
        \right],
        \end{align}
\end{subequations}
with the functions $\zeta_{\pm}^{\alpha}$ e $\xi_{\pm}^{\alpha}$ defined as
\begin{subequations}
        \begin{align}
        \zeta_{+}^{\alpha}(t) &=  T_{+}^{\alpha}(t) - \lambda_{\alpha}^{\ast}(0) W_{-}^{\alpha}(t),
        \\
        \zeta_{-}^{\alpha}(t) &= \lambda_{\alpha}(0) T_{+}^{\alpha}(t) - [\Lambda_{\alpha}(0) + |\lambda_{\alpha}(0)|^{2}] W_{-}^{\alpha}(t),
        \\
        \xi_{+}^{\alpha}(t) &=  W_{+}^{\alpha}(t) - \lambda_{\alpha}^{\ast}(0) T_{-}^{\alpha}(t) ,
        \\
        \xi_{-}^{\alpha}(t) &= \lambda_{\alpha}(0) W_{+}^{\alpha}(t) -  [\Lambda_{\alpha}(0) + |\lambda_{\alpha}(0)|^{2}] T_{-}^{\alpha}(t),
        \end{align}
\end{subequations}
in terms of the functions $W_{\pm}^{\alpha}(t)$ and $T_{\pm}^{\alpha}(t)$ arising from Eq. \eqref{WT}. Admitting $\hat{u}_{\alpha}(t)$ as a unitary operator, it must necessarily belong to the Lie group $U(2)$, and this fact implies that
\begin{equation}
|\mathrm{det}\,\hat {u}_{\alpha}(t)|^{2} = 1.
\end{equation}
From Eq. \eqref{uU}, the determinant of $\hat{u}_{\alpha}(t)$ can be
determined as follows
\begin{align}
\mathrm{det}\,\hat{u}_{\alpha}(t)
&= \mathrm{det}\,[\hat{\eta}_{\alpha}(t)\hat{U}_{\alpha}(t)\hat{\eta}_{\alpha}^{-1}(0)]
\nonumber\\
&=e^{2[\kappa_{\alpha}(t)-\kappa_{\alpha}(0)]}D_{\alpha}(t)
\nonumber\\
&=e^{2[\kappa_{\alpha}(t)-\kappa_{\alpha}(0)]}e^{\mathrm{Re}[\ln D_{\alpha}(t)]} e^{i\mathrm{Im}[\ln D_{\alpha}(t)]},
\end{align}
in which $D_{\alpha}(t)\equiv\mathrm{det}\,\hat{U}_{\alpha}(t)$ is
written in terms of the $
\mathcal{C}_{\alpha}\left(\Delta_{\alpha}t^{\alpha}\right)$ and
$\mathcal{S}_{\alpha}\left(\Delta_{\alpha}t^{\alpha}\right)$ functions
as
\begin{align}
D_{\alpha}(t)
  = {}
  &
W_{+}^{\alpha}(t)W_{-}^{\alpha} - T_{+}^{\alpha}(t)T_{-}^{\alpha}
\nonumber\\
  = {}
  &
\mathcal{C}^{2}_{\alpha}\left(\Delta_{\alpha}t^{\alpha}\right)
-
(-1)^{-\alpha}\mathcal{S}^{2}_{\alpha}\left(\Delta_{\alpha}t^{\alpha}\right).
\end{align}
Since the conditions $|\mathrm{det}\,\hat {u}_{\alpha}(t)|^{2} = 1$ must
be verified, it implies to the Dyson map parameter $\kappa_{\alpha}(t)$
to have the form 
\begin{equation}
\kappa_{\alpha}(t) =\kappa_{\alpha}(0) -\frac{1}{2}\mathrm{Re}[\ln{D_{\alpha}(t)}],
\end{equation}
in such a way that the determinant is an imaginary phase written as
\begin{equation}
\mathrm{det}\,\hat{u}_{\alpha}(t) = e^{i\mathrm{Im}[\ln{D_{\alpha}(t)}] }.
\end{equation}

The other parameters of the Dyson map can be obtained by imposing the unitarity condition
\begin{equation}
\hat{u}_{\alpha}^{-1}(t)=\hat{u}_{\alpha}^{\dagger}(t),
\end{equation}
that leads to the following relation among the coefficients of matrix \eqref{u_matrix}:
\begin{align}
\tau_{\pm}^{\alpha}(t) &=\mp e^{i\mathrm{Im}[\ln{D_{\alpha}(t)}] }[\varpi_{\mp}^{\alpha}(t)]^{\ast}.
\end{align}
From these equations, we can obtain the exact forms of $\lambda_{\alpha}$ and $\Lambda_{\alpha}$:
\begin{subequations}
        \begin{align}
          \lambda_{\alpha} (t) = {}
          &
            - \frac{\xi_{+}^{\alpha}(\zeta_{+}^{\alpha})^{\ast} +
            \xi_{-}^{\alpha}(\zeta_{-}^{\alpha})^{\ast}}{|\xi_{+}^{\alpha}|^{2}
            +
            |\xi_{-}^{\alpha}|^{2}
            + \Lambda_{\alpha}(0)e^{\mathrm{Re}[\ln{D_{\alpha}}]}},
        \\
          \Lambda_{\alpha}(t) = {}
          &
         \frac{|\zeta_{+}^{\alpha}|^{2} + |\zeta_{-}^{\alpha}|^{2} + \Lambda_{\alpha}(0)e^{\mathrm{Re}[\ln{D_{\alpha}}]}}{ |\xi_{+}^{\alpha}|^{2} + |\xi_{-}^{\alpha}|^{2} + \Lambda_{\alpha}(0)e^{\mathrm{Re}[\ln{D_{\alpha}}]}}
         - |\lambda_{\alpha}|^{2},
        \end{align}
\end{subequations}
where these forms are obtained by considering the equality
\begin{equation}\label{cdt}
[\zeta_{+}^{\alpha}\xi_{-}^{\alpha} - \zeta_{-}^{\alpha}\xi_{+}^{\alpha}]e^{-i\mathrm{Im}[\ln{D_{\alpha}}]} = \Lambda_{\alpha}(0)e^{\mathrm{Re}[\ln{D_{\alpha}}]}.
\end{equation}
Notice that these functions depend only on the fractional-time evolution
parameters and on the initial values of Dyson map parameters due to the
functions $\xi_{\pm}^{\alpha}$ and $\zeta_{\pm}^{\alpha}$.
Therefore, the matrix corresponding to the unitary time-evolution
operator can be rewritten as
\begin{equation}
\label{u_matrixF}
\hat{u}_{\alpha}(t)
=
e^{\frac{i}{2}\mathrm{Im}[\ln{D_{\alpha}(t)}] }
\left[
\begin{matrix}
\varpi^{\alpha}(t) & \tau^{\alpha}(t) \\
-[\tau^{\alpha}(t)]^{\ast} & [\varpi^{\alpha}(t)]^{\ast}
\end{matrix}
\right],
\end{equation}
with
\begin{subequations}
  \begin{align}
    \varpi^{\alpha}(t) = {}
    &
      e^{-\frac{i}{2}\mathrm{Im}[\ln{D_{\alpha}}] }\varpi_{+}^{\alpha}= e^{\frac{i}{2}\mathrm{Im}[\ln{D_{\alpha}}] }(\tau_{-}^{\alpha})^{\ast},
        \\
    \tau^{\alpha}(t)   = {}
    &
      e^{-\frac{i}{2}\mathrm{Im}[\ln{D_{\alpha}}] }\varpi_{-}^{\alpha}= -e^{\frac{i}{2}\mathrm{Im}[\ln{D_{\alpha}}] }(\tau_{+}^{\alpha})^{\ast},
        \end{align}
\end{subequations}
which satisfy the relation
\begin{equation}
|\varpi^{\alpha}(t)|^{2} + |\tau^{\alpha}(t)|^{2} = 1.
\end{equation}
In this manner, we obtain the Dyson map associated with the metric
operator, about which the fractional dynamics is unitary 
\begin{widetext}
\begin{align}
\langle \Psi^{\alpha}(t)\vert \Psi^{\alpha}(t)\rangle_{\Theta_{\alpha} (t)}
=
\langle \Psi^{\alpha}(0)\vert \Psi^{\alpha}(0)\rangle_{\Theta_{\alpha} (0)}
=
\langle \psi^{\alpha}(0) \vert \psi^{\alpha}(0)\rangle
=
\langle \psi^{\alpha}(t) \vert \psi^{\alpha}(t)\rangle.
\end{align}
\end{widetext}
This relation reflects that the probability conservation in the
fractional scenario can be achieved by defining a suitable
time-dependent metric through the non-Hermitian formalism.
Notice that the explicit form of the Hermitian and non-Hermitian
Schr\"odinger equations are not necessary in this case:
in the models considered here, the solutions of fractional Schr\"odinger
equations are well known, and the main task boils down to finding the
form of the metric operator.
\begin{figure*}[!t]
	\begin{center}
		\includegraphics[width=0.45\linewidth]{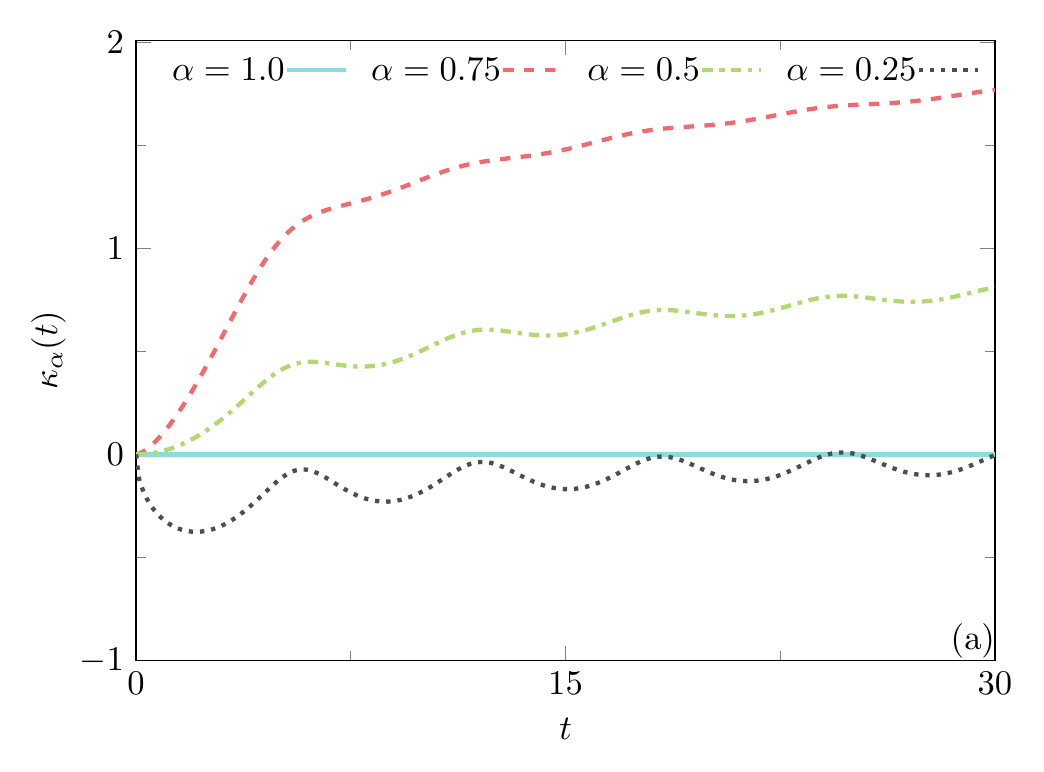}
		\includegraphics[width=0.45\linewidth]{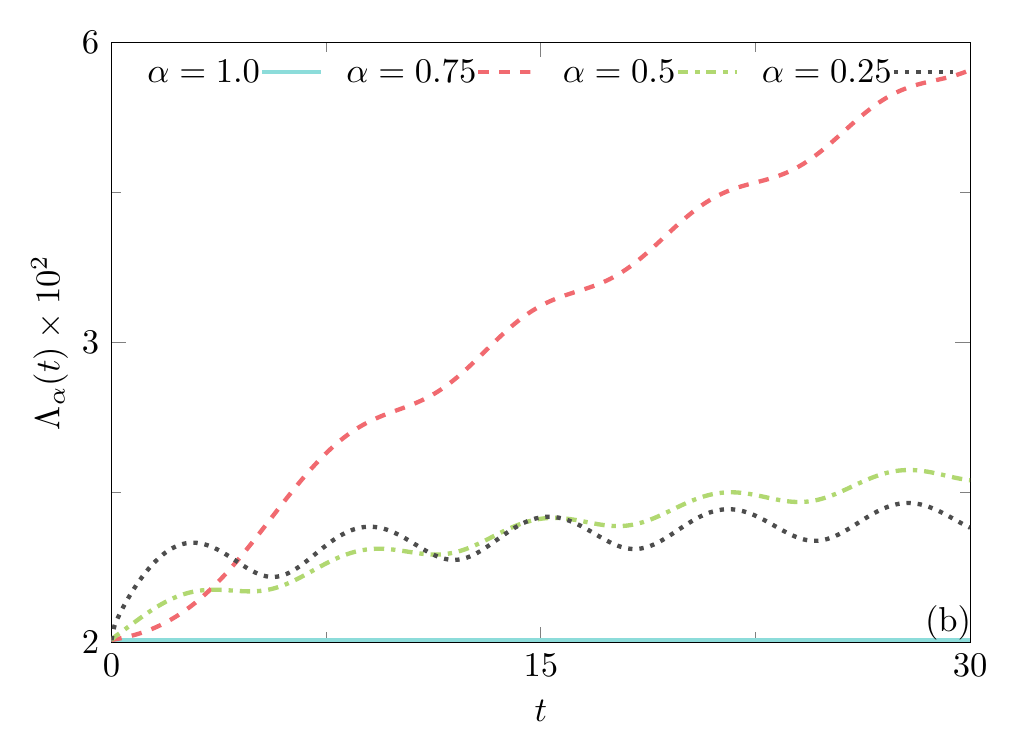}
		\\
		\includegraphics[width=0.45\linewidth]{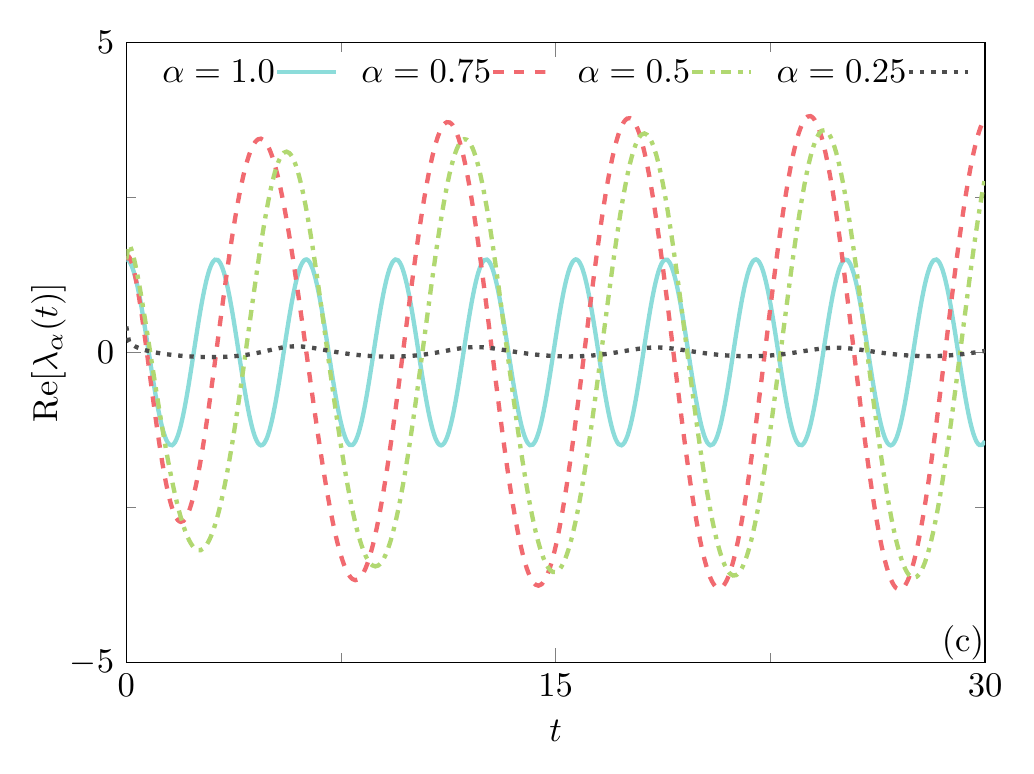}
		\includegraphics[width=0.45\linewidth]{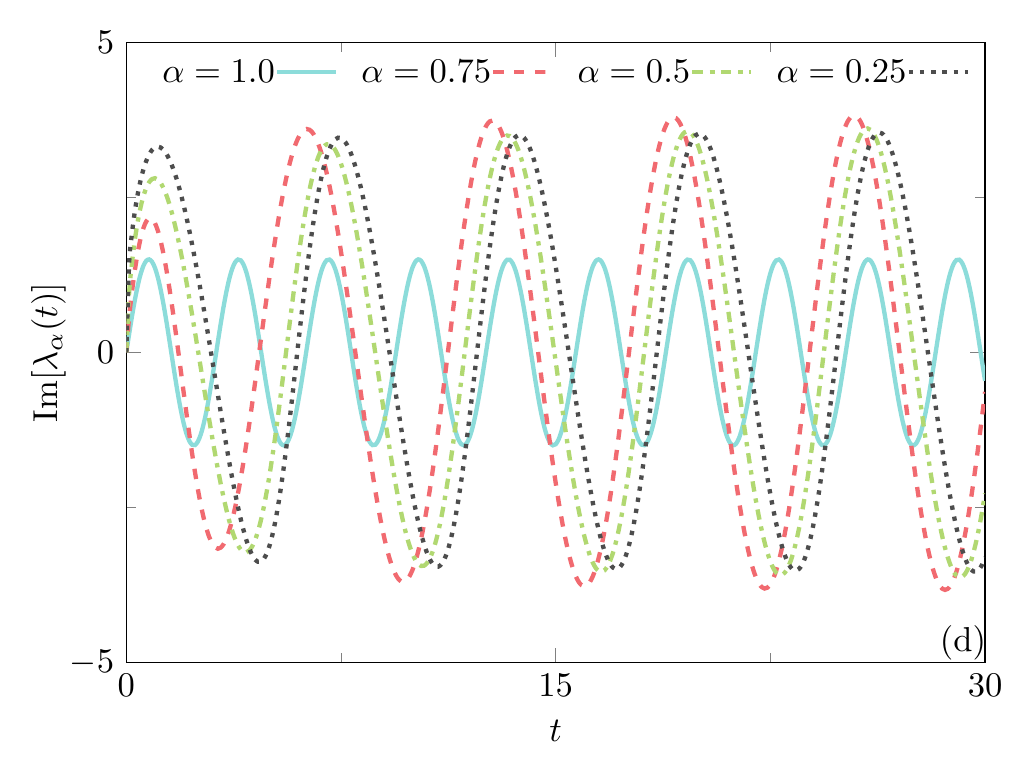}
	\end{center}
	\caption{
		\label{fig:EX1DysonMap}
		Time-evolution of the Dyson map parameters $\kappa_{\alpha}$,
		$\epsilon_{\alpha}$, $\mu_{\alpha}$ are plotted for $\alpha=1.0$
		(solid line), $\alpha=0.75$ (dashed line), $\alpha=0.5$ (dot-dashed
		line) and $\alpha=0.25$ (dotted line). 
		We set $\Delta_{\alpha}=1\,\mathtt{s}^{-\alpha}$, and the initial
		conditions: $\kappa_{\alpha}(0)=0$, $\Lambda_{\alpha}(0)=2$ and
		$|\lambda_{\alpha}(0)|=3/2$.}
\end{figure*}

\section{Applications}
\label{sec:Appl}
\subsection{Spin-$1/2$ interacting with a longitudinal magnetic field}
\label{subsec:1}

As a first example, we consider a nuclei with spin-$1/2$ placed in a
longitudinal magnetic field $\textbf{B}^{\alpha} = (0,0,B_0^{\alpha})$.
The interaction between the nuclei and the external magnetic field is
described by the Zeeman Hamiltonian
\begin{equation}
\label{HZeeman}
    \hat{\mathcal{H}}_{0}^{\alpha}
    =
    -\hat{\boldsymbol{\mu}}_{\text{I}}^{\alpha}\cdot\mathbf{B}^{\alpha}
    =
    -\frac{\hbar_{\alpha}\omega_{\text{L}}^{\alpha}}{2}\hat{\sigma}_{3},
\end{equation}
with the magnetic moment of the nuclei defined as
$\hat{\boldsymbol{\mu}}_{I}^{\alpha}=
\hbar_{\alpha}\gamma_{\alpha}\hat{\boldsymbol{\sigma}}/2$
where  $\gamma_{\alpha}$ is the gyromagnetic ratio and
$\omega_{\text{L}}^{\alpha}=\gamma_{\alpha} B_{0}^{\alpha}$, the Larmor
frequency of the nuclei.
For this model,  the Hamiltonian coefficients in Eq. \eqref{H0} are
given by $\omega_{1}^{\alpha}=\omega_{2}^{\alpha}=0$ and
$\omega_{3}^{\alpha} = -\omega_{\text{L}}^{\alpha}/2$, which leads to
$\Delta_{\alpha}=\omega_{\text{L}}^{\alpha}/2$. 
Based on the previous results, we plot the time-dependent Dyson map
parameters $\kappa_{\alpha}, \Lambda_{\alpha},\lambda_{\alpha}$ in
Fig. \ref{fig:EX1DysonMap}, assuming the initial values
$\kappa_{\alpha}(0)=0$, $\Lambda_{\alpha}(0)=2$ and
$|\lambda_{\alpha}(0)|=3/2$.

\begin{figure*}[!ht]
  \begin{center}
     \includegraphics[width=0.45\linewidth]{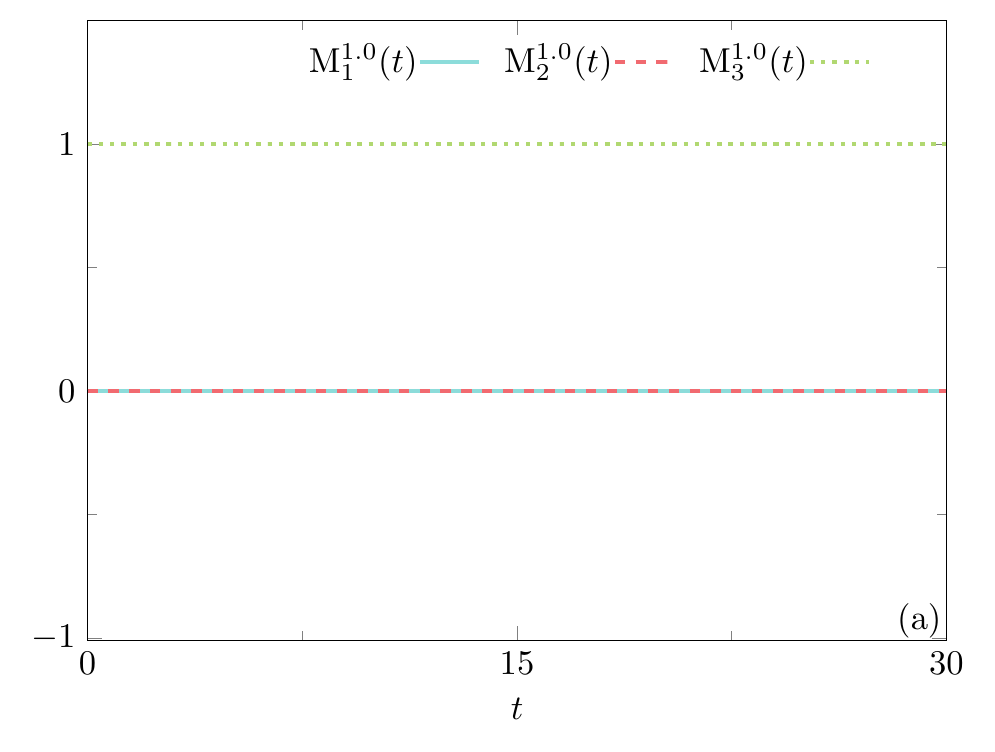}
     \includegraphics[width=0.45\linewidth]{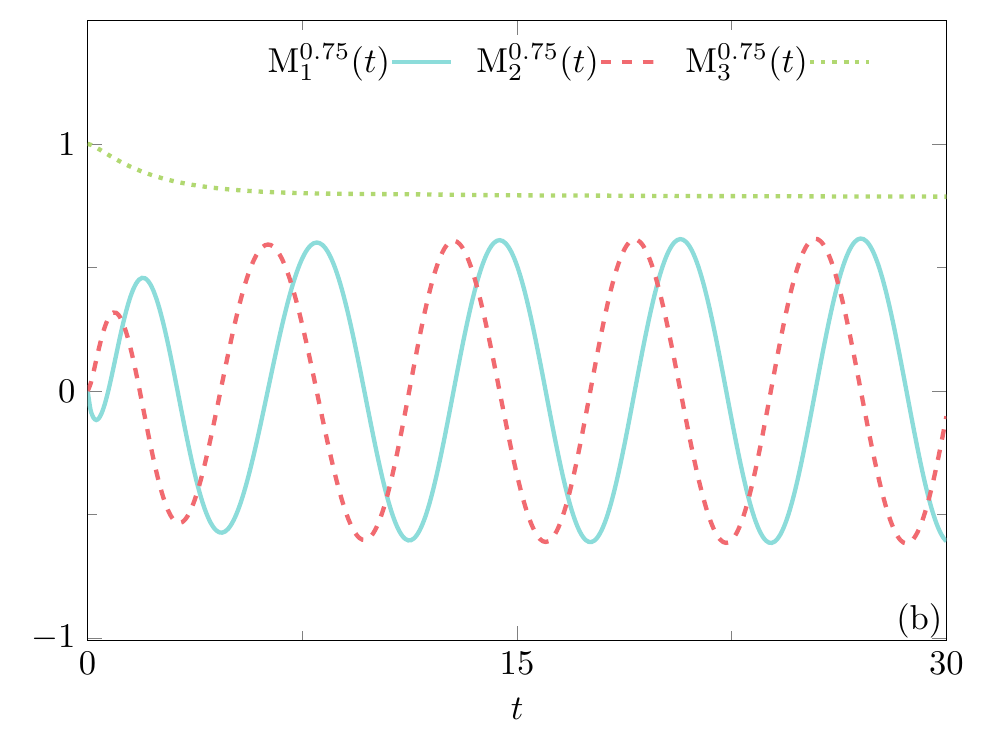}
     \\
     \includegraphics[width=0.45\linewidth]{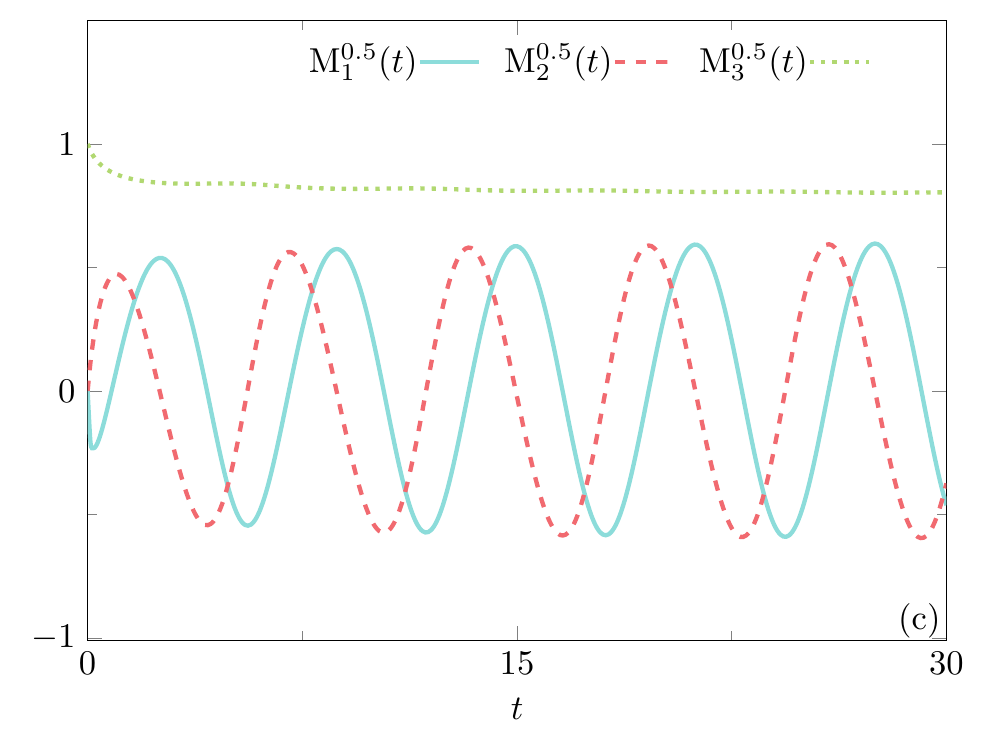}
     \includegraphics[width=0.45\linewidth]{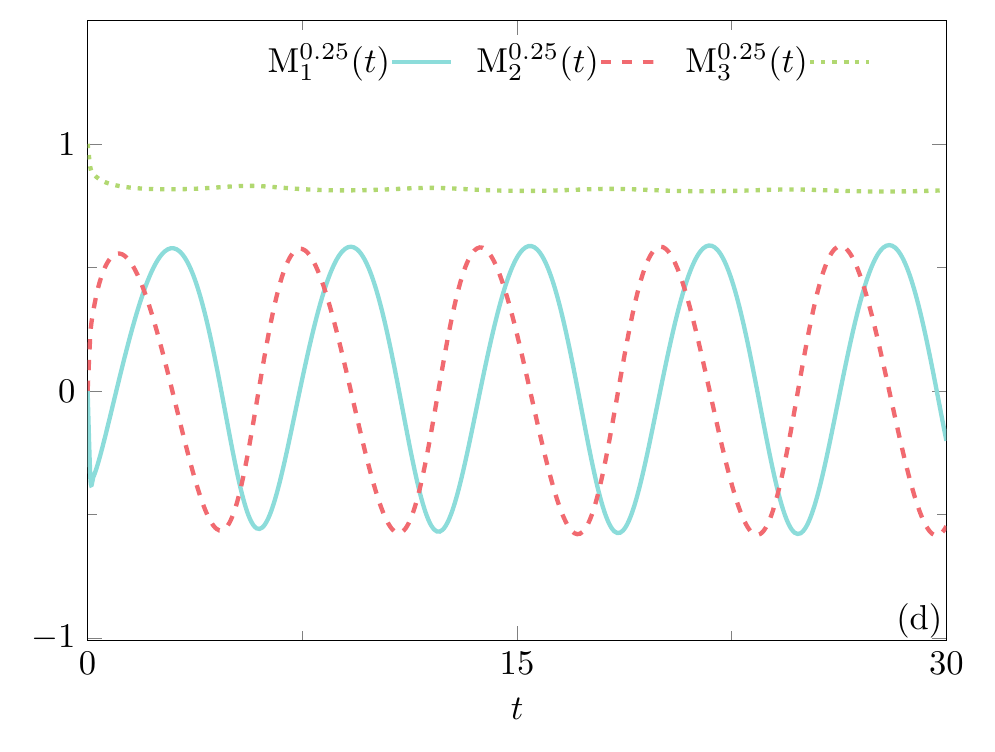}
  \end{center}
\caption{
  \label{fig:EX1Magnetizations}
  Starting from the spin-up state, we plot the time-evolution of the
  dimensionless magnetization components $\mathrm{M}_{k}(t)$ of a
  spin-$1/2$ nuclei in the presence of a longitudinal magnetic field for
  the cases: (a) $\alpha=1.0$ (b) $\alpha=0.75$ (c) $\alpha=0.5$ (d)
  $\alpha=0.25$.}
\end{figure*}

We apply the unitary time-evolution \eqref{u_matrixF} to compute the
three components of the dimensionless normalized magnetization,
$\mathrm{M}^{\alpha}_{k}(t)$ ($k=1,2,3$), for one-half spin system,
which are determined as 
\begin{equation}
\label{Magnetization}
    \mathrm{M}^{\alpha}_{k}(t) = \frac{\langle \psi^{\alpha}(t) \vert \hat{\sigma}_{k} \vert \psi^{\alpha}(t) \rangle}{\langle \psi^{\alpha}(t) \vert \psi^{\alpha}(t) \rangle }\,,
\end{equation}
where the quantum state $\vert \psi^{\alpha}(t) \rangle$ is given by Eq. \eqref{UFS1}.
Starting from the spin-up state that corresponds to $\vert \psi^{\alpha}
(0)\rangle = (1, 0)^{\mathrm{T}}$, we obtain the magnetization
components in terms of the functions $\varpi^{\alpha}(t)$ and
$\tau^{\alpha}(t)$ in the following forms
\begin{subequations}
        \label{Ex1Magnetizations}
        \begin{align}
        \mathrm{M}^{\alpha}_{1}(t) &=
        -2\mathrm{Re}[\varpi^{\alpha}(t)\tau^{\alpha}(t)],
        \\
        \mathrm{M}^{\alpha}_{2}(t) &=
        2\mathrm{Im}[\varpi^{\alpha}(t)\tau^{\alpha}(t)],
        \\
        \mathrm{M}^{\alpha}_{3}(t) &=
        |\varpi^{\alpha}(t)|^{2} - |\tau^{\alpha}(t)|^{2}.
        \end{align}
\end{subequations}
These magnetization components are evaluated for different values of
$\alpha$ in Fig. \ref{fig:EX1Magnetizations} assuming
$\Delta_{\alpha}=1\,\mathtt{s}^{-\alpha}$.
Figure \ref{fig:EX1Magnetizations}(a) corresponds to the case where
$\alpha=1.0$, from where we can observe that the quantum state remains at
the spin-up state as time goes by.
This is an expected result since the initial state corresponds to the
eigenstate of the Hermitian Hamiltonian \eqref{HZeeman}. 
In the cases where $\alpha\neq 1.0$, the Dyson map introduces transverse
fields that perturb the initial state of the system, applying a torque
that rotates the nuclear magnetic moment around the effective magnetic
field.
Sounds interesting in this model the fact that this behavior is slightly
modified by changes in the $\alpha$ parameter.
The effect of the fractional parameter $\alpha$ is a short-lived burst
of energy in the initial configuration caused by a remarkable change of
state, which makes the magnetization components start to oscillate
around the $x_{3}$-axis.

\subsection{One-site Lee-Yang Chain}
Another model of physical interest considered in our approach is the
discrete lattice version of the Yang-Lee model, which was proposed in
Ref. \cite{gehlen:91}.
It is described as an Ising spin chain in the presence of a longitudinal
magnetic field in the $x_{3}$-direction and of a transverse pure
imaginary field in the $x_{1}$-direction.
The correspondent total Hamiltonian operator reads as
\begin{equation}
\label{YAN}
\hat{\mathcal{H}}_{0}^{\alpha}=
-\frac{\hbar_{\alpha}}{2}\sum_{i=1}^{N}\left(\hat{\sigma}_{3}^{(i)}
  + J_{\alpha} \hat{\sigma}_{1}^{(i)}\hat{\sigma}_{1}^{(i+1)}
  + i\xi_{\alpha}\hat{\sigma}_{1}^{(i)}  \right),
\end{equation}
where $J_{\alpha}$ is the coupling constant with $\xi_{\alpha}$
proportional to the intensity of imaginary field in $x_{1}$-direction.
Also, the periodic condition
$\hat{\sigma}_{1}^{(N+1)}=\hat{\sigma}_{1}^{(1)}$ is assumed in this
case.

For our purposes, we consider the particular case of a one-site spin chain ($N=1$), and take the set of the coupling constant to be $J_{\alpha}=0$. Under these assumptions, the non-Hermitian Hamiltonian operator \eqref{YAN} reduces to the form,
\begin{equation}
\label{XX}
   \hat{\mathcal{H}}_{0}^{\alpha}=-\frac{\hbar_{\alpha}}{2}\hat{\sigma}_{3} -i\frac{\hbar_{\alpha}\xi_{\alpha}}{2}\hat{\sigma}_{1}.
\end{equation}
In \eqref{XX} the single spin interacts with a real field in the
$x_{3}$-direction and with a pure imaginary field in the
$x_{1}$-direction in which the intensity is proportional to the real
parameter $\xi_{\alpha}$.
Moreover, the Hamiltonian coefficients in Eq. \eqref{H0} take on the
following values:  $\omega_{1}^{\alpha}=-i\xi_{\alpha}/2$,
$\omega_{2}^{\alpha}=0$ and $\omega_{3}^{\alpha} =
-1/2\,\mathtt{s}^{-\alpha}$. 

\begin{figure*}[!ht]
  \begin{center}
    \includegraphics[width=0.45\linewidth]{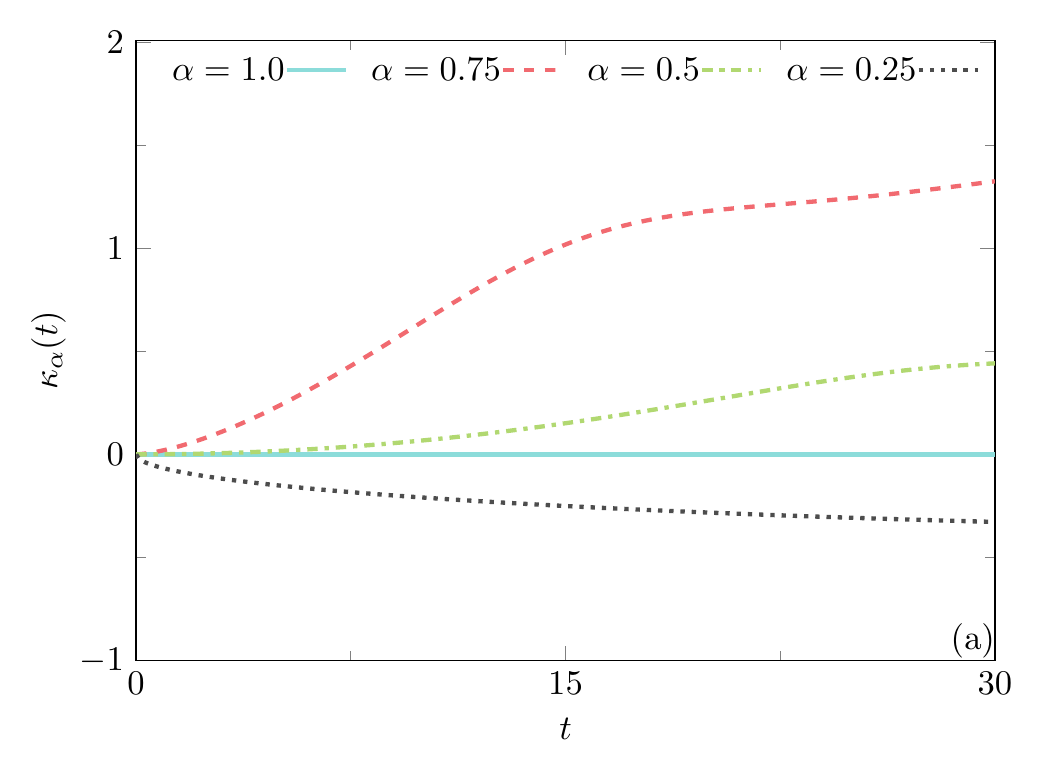}
    \includegraphics[width=0.45\linewidth]{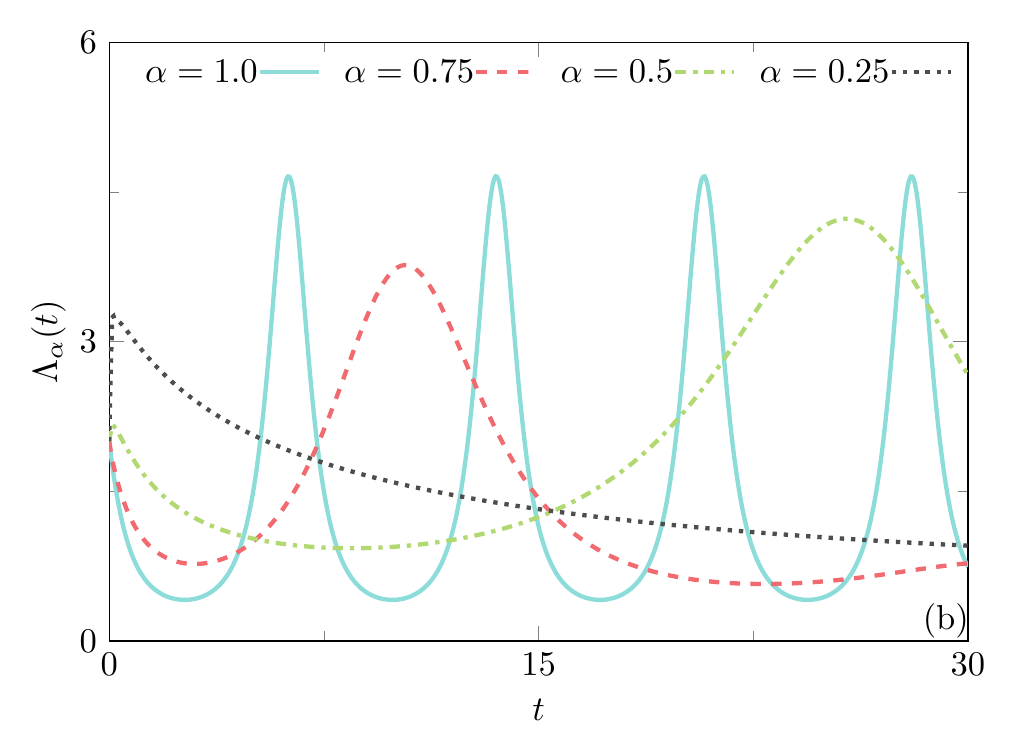} \\
    \includegraphics[width=0.45\linewidth]{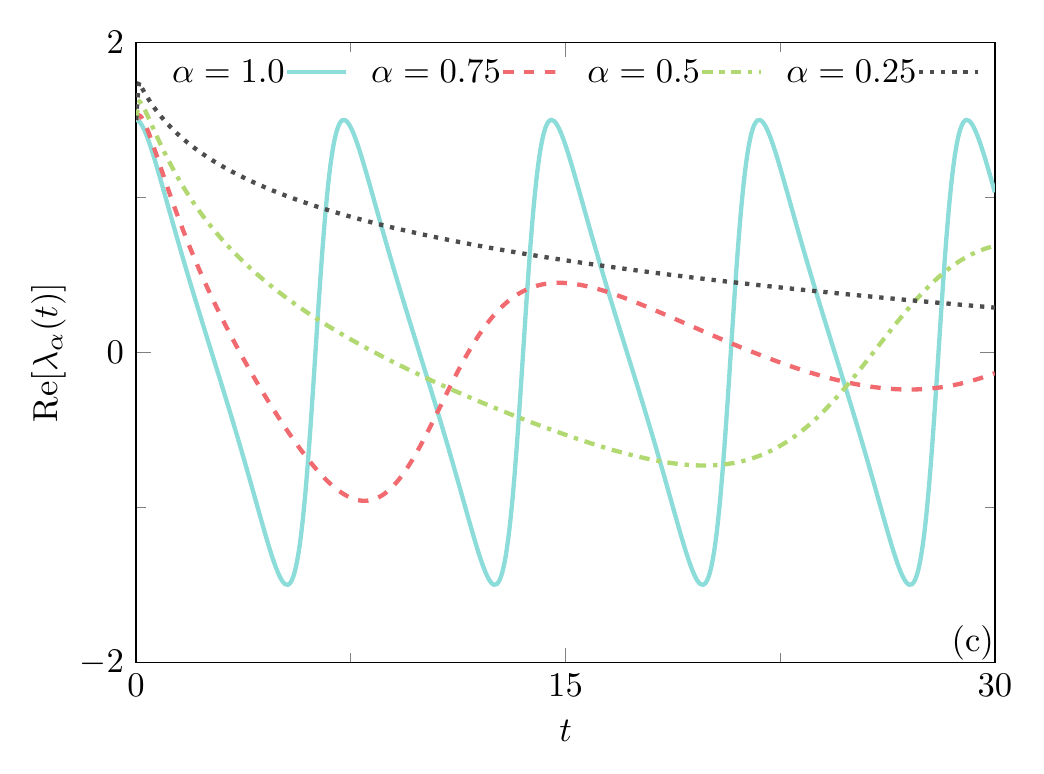}
    \includegraphics[width=0.45\linewidth]{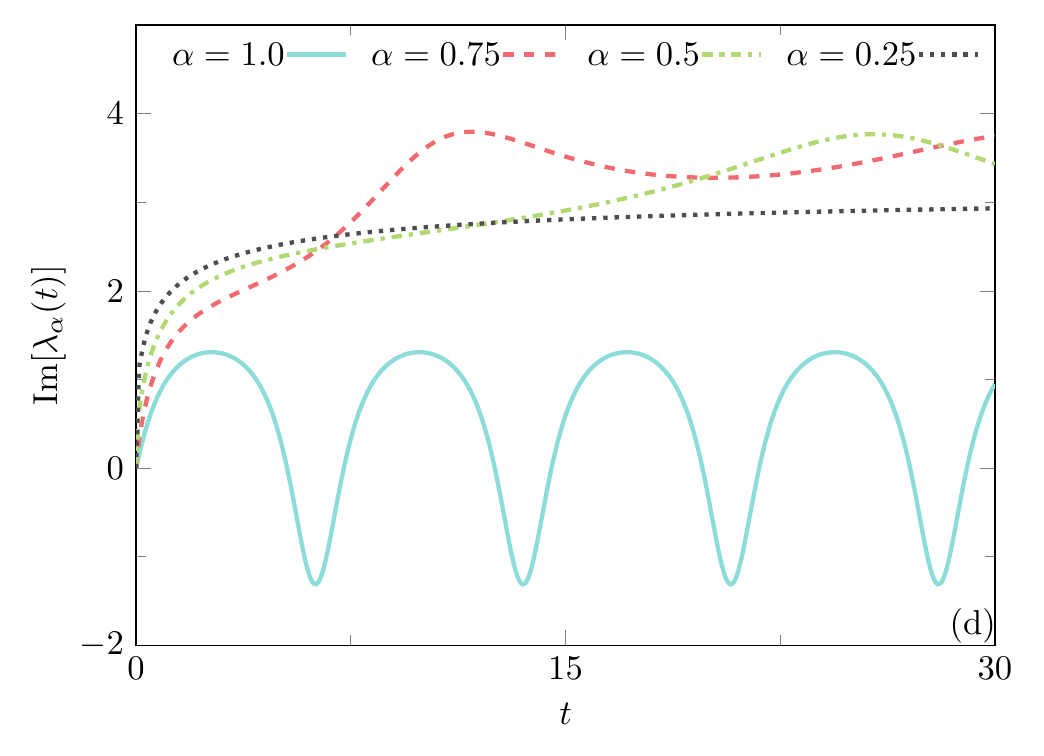}
  \end{center}
  \caption{
    \label{fig:EX2DysonMap}
    The time-evolution of the Dyson map parameters $\kappa_{\alpha}$,
    $\epsilon_{\alpha}$, $\mu_{\alpha}$ are plotted for $\alpha=1.0$ (solid
    line), $\alpha=0.75$ (dashed line), $\alpha=0.5$ (dot-dashed line) and
    $\alpha=0.25$ (dotted line). 
    We set $\Delta_{\alpha}= \sqrt{3}/4\,\mathtt{s}^{-\alpha}$
    ($\omega_{3}=-1/2\,\mathtt{s}^{-\alpha}$ and
    $\xi_{\alpha}=1/2\,\mathtt{s}^{-\alpha}$), and the initial
    conditions: $\kappa_{\alpha}(0)=0$, $\Lambda_{\alpha}(0)=2$ and
    $|\lambda_{\alpha}(0)|=3/2$. 
  }
\end{figure*}

Different from the first application in which the Hamiltonian operator
is Hermitian, we now start from a non-Hermitian one in the
fractional-time scenario.
The non-unitarity of the time-evolution comes from both the
non-hermiticity due to the imaginary field in $x_{1}$-direction and the
fractional order time-derivative.
Therefore, the Dyson map considers both properties to provide a unitary
time-evolution operator.
The time-dependent parameters of Dyson map are plotted in
Fig. \ref{fig:EX2DysonMap} taking into account distinct values of
$\alpha$ under the same initial conditions assumed in the first
application model:
$\xi_{\alpha} = 1/2\,\mathtt{s}^{-\alpha}$, and consequently
$\Delta_{\alpha} = \sqrt{3}/4\,\mathtt{s}^{-\alpha}$.

\begin{figure}[!h]
	\begin{center}
		\includegraphics[width=0.912\linewidth]{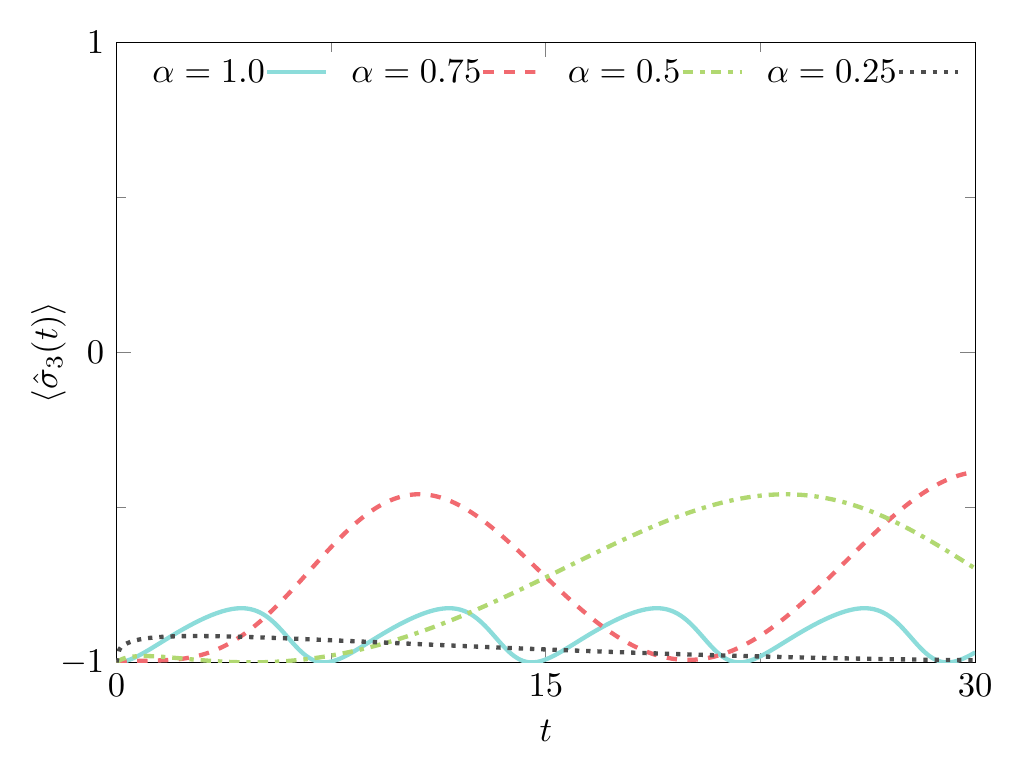}
	\end{center}
	\caption{ \label{fig:EX2Population}
		Starting from the spin-down state, we plot the difference of
		population $\langle\hat{\sigma}_{3}(t)\rangle$ between the two spin
		states  for the cases: $\alpha=1.0$ (solid), $\alpha=0.75$ (dashed),
		$\alpha=0.5$ (dot-dashed) and $\alpha=0.25$ (dotted).} 
\end{figure}

The unitary time-evolution arising from the general form
\eqref{u_matrixF} for this system allows us to determine the dynamics of
the population difference between the two level states, namely, spin-up
and spin-down, which are represented by the state vectors
$(1,0)^{\mathrm{T}}$ and $(0,1)^{\mathrm{T}}$, respectively. 
We assume the system starts from the spin-down state represented by
$\vert \psi^{\alpha}(0) \rangle = (0,1)^{\mathrm{T}}$ for which the
population difference evolves in time in according to 
\begin{equation}\label{Dp}
\langle\hat{\sigma}_{3}(t)\rangle = \frac{\langle \psi^{\alpha}(t) \vert \hat{\sigma}_{3}\vert \psi^{\alpha}(t) \rangle}{\langle \psi^{\alpha}(t)\vert \psi^{\alpha}(t) \rangle} = |\tau^{\alpha}(t)|^{2} - |\varpi^{\alpha}(t)|^{2}.
\end{equation}
In Fig. \ref{fig:EX2Population}  we show the behavior of the population
difference \eqref{Dp} for distinct values of $\alpha$.
For $\alpha=1.0$, the population difference oscillates with a small
amplitude near $-1$, which means that most of the population remains in
the spin-down state, and there is a tiny population in the spin-up
state.
Most of the population occupies the spin-up state for $\alpha=0.75$ and
$\alpha=0.5$.
Therefore, the population difference oscillates from the $-1$ to close
to $-1/2$ as time goes by:  the population difference oscillates faster
for $\alpha=0.75$ than $\alpha=0.5$.
When $\alpha=0.25$, a minimal change in populations at the spin-down
state occurs, but it returns to $-1$ during the observed time interval.

\begin{figure*}[t]
	\begin{center}
		\includegraphics[width=0.45\linewidth]{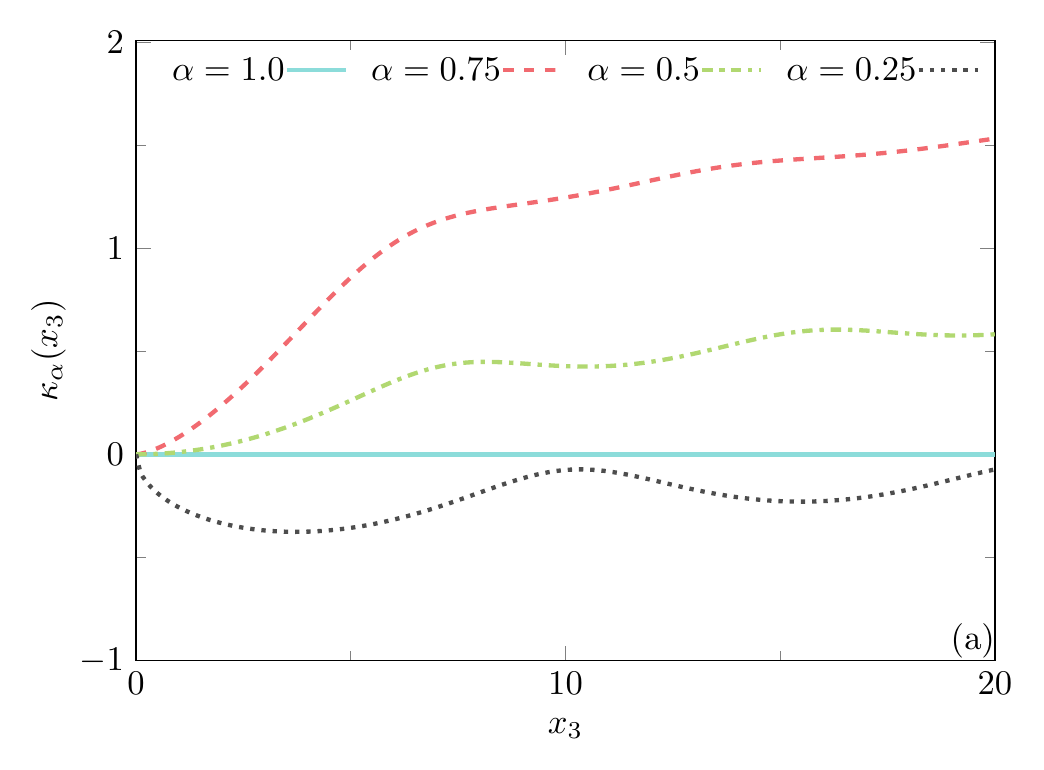}
		\includegraphics[width=0.45\linewidth]{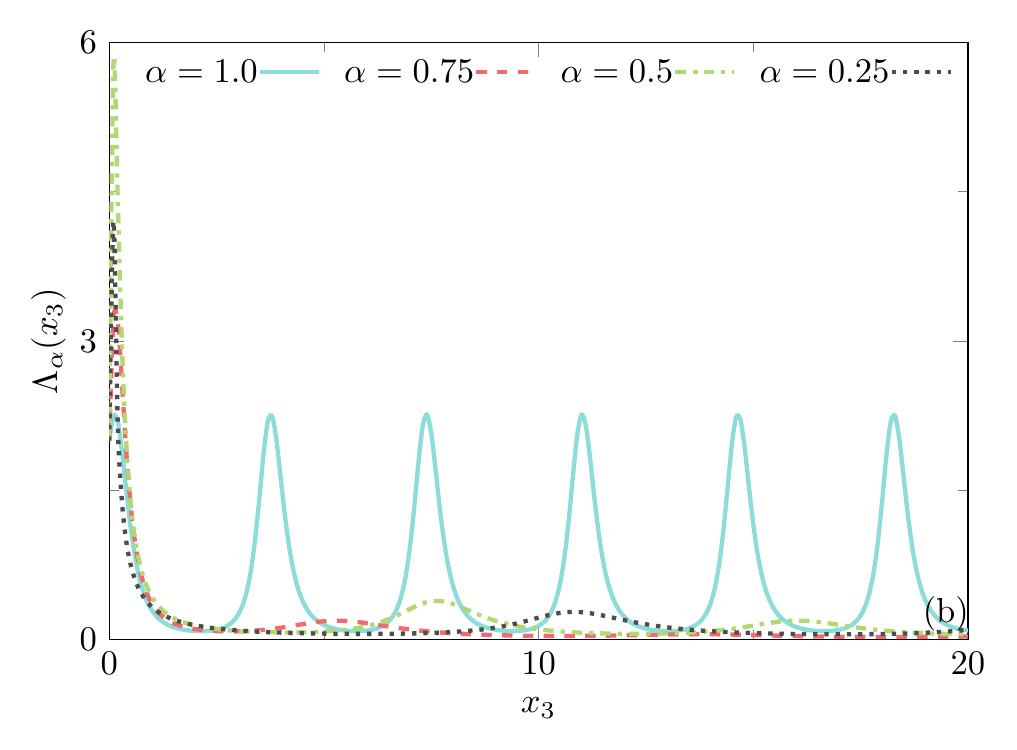}
		\\
		\includegraphics[width=0.45\linewidth]{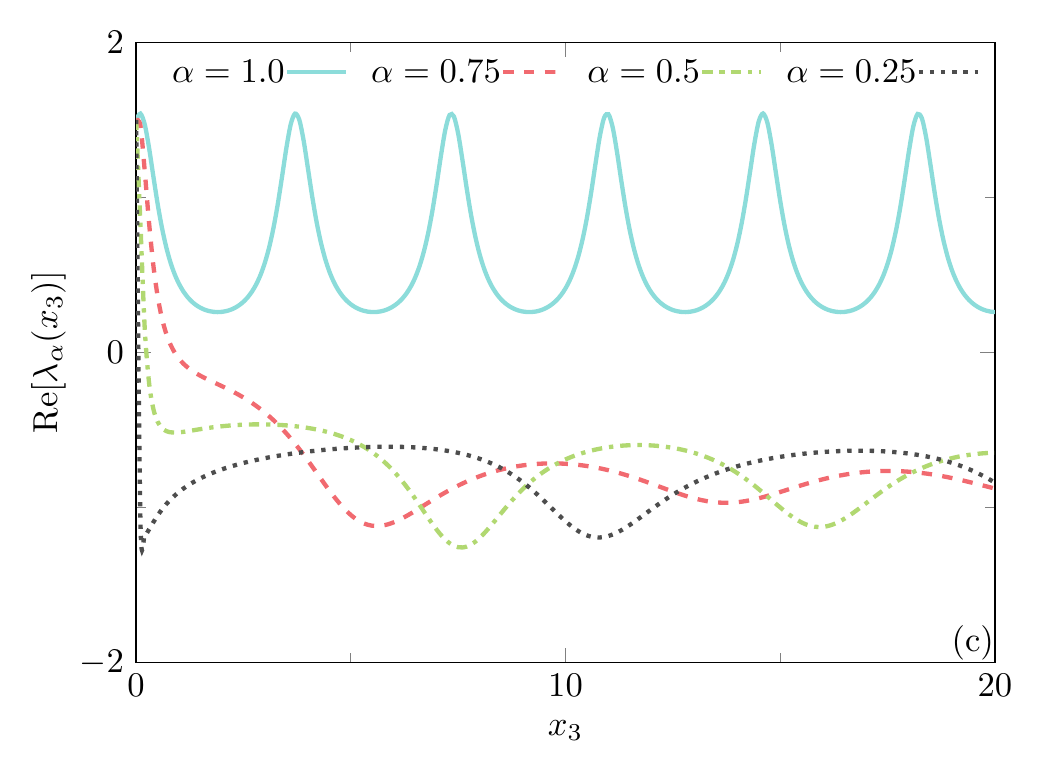}
		\includegraphics[width=0.45\linewidth]{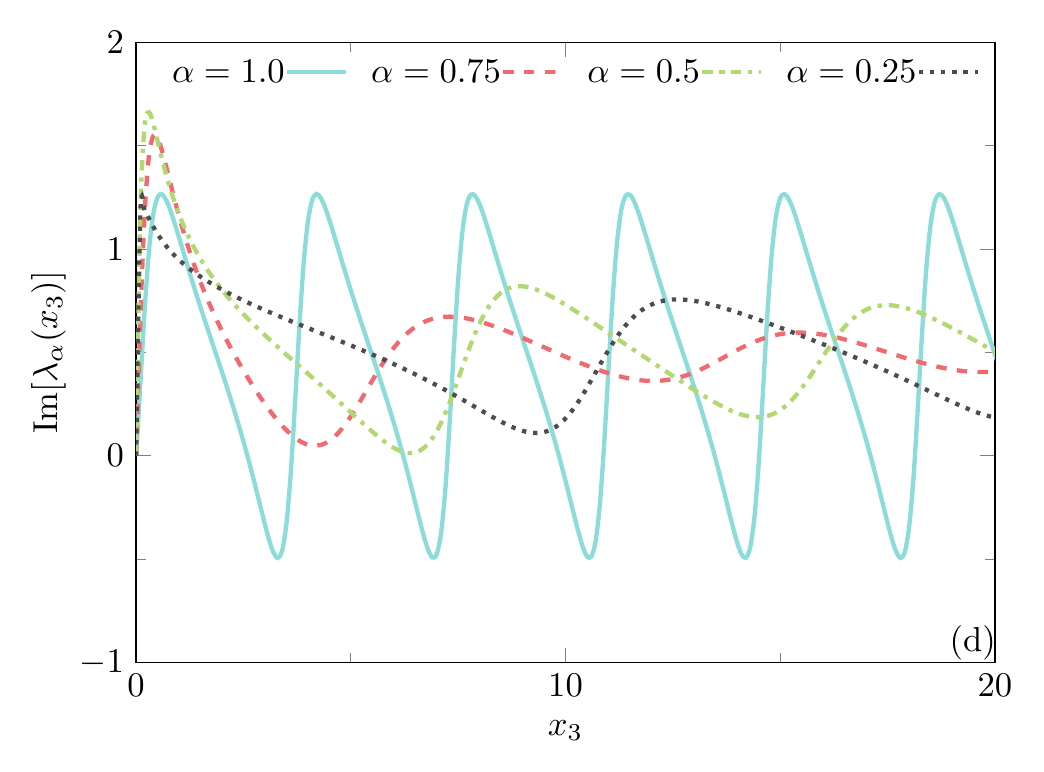}
	\end{center}
	\caption{
		\label{fig:EX3DysonMap}
		The time-evolution of the Dyson map parameters $\kappa_{\alpha}$,
		$\epsilon_{\alpha}$, $\mu_{\alpha}$ are plotted for $\alpha=1.0$
		(solid line), $\alpha=0.75$ (dashed line), $\alpha=0.5$ (dot-dashed
		line) and  $\alpha=0.25$ (dotted line).
		We set $\Delta_{\alpha}=\sqrt{3}/2$ ($\varsigma_{\alpha}=1$ and
		$\varepsilon_{\alpha}=1/2$), and the initial conditions:
		$\kappa_{\alpha}(0)=0$, $\Lambda_{\alpha}(0)= 2$, and
		$|\lambda_{\alpha}(0)|=3/2$.
	}
\end{figure*}

\begin{figure*}[!t]
	\begin{center}
		\includegraphics[width=0.45\linewidth]{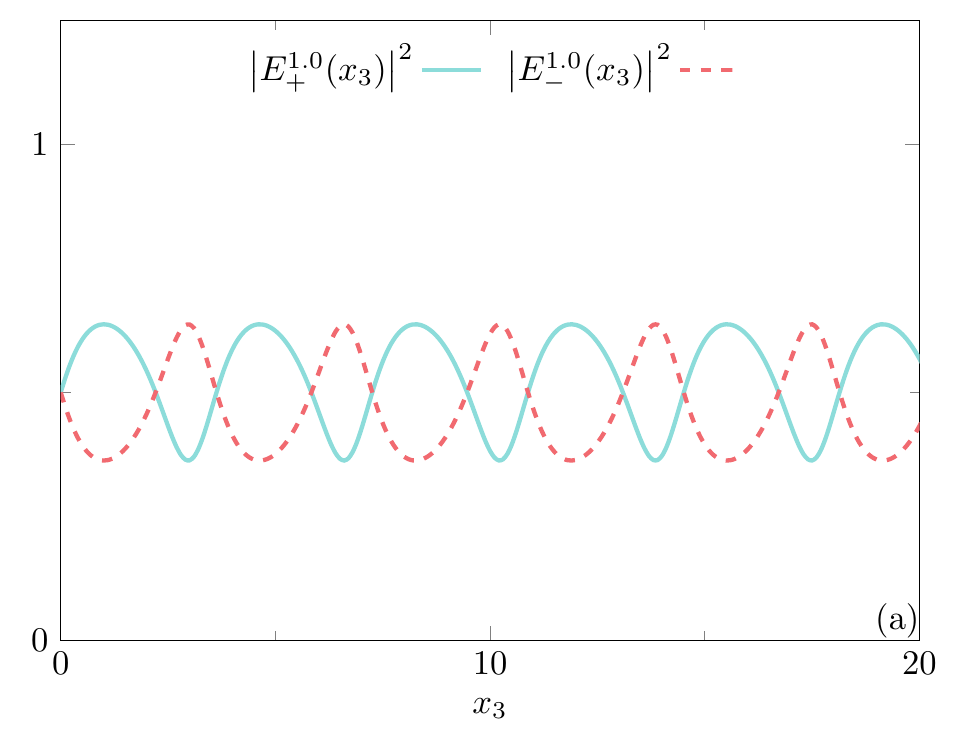}
		\includegraphics[width=0.45\linewidth]{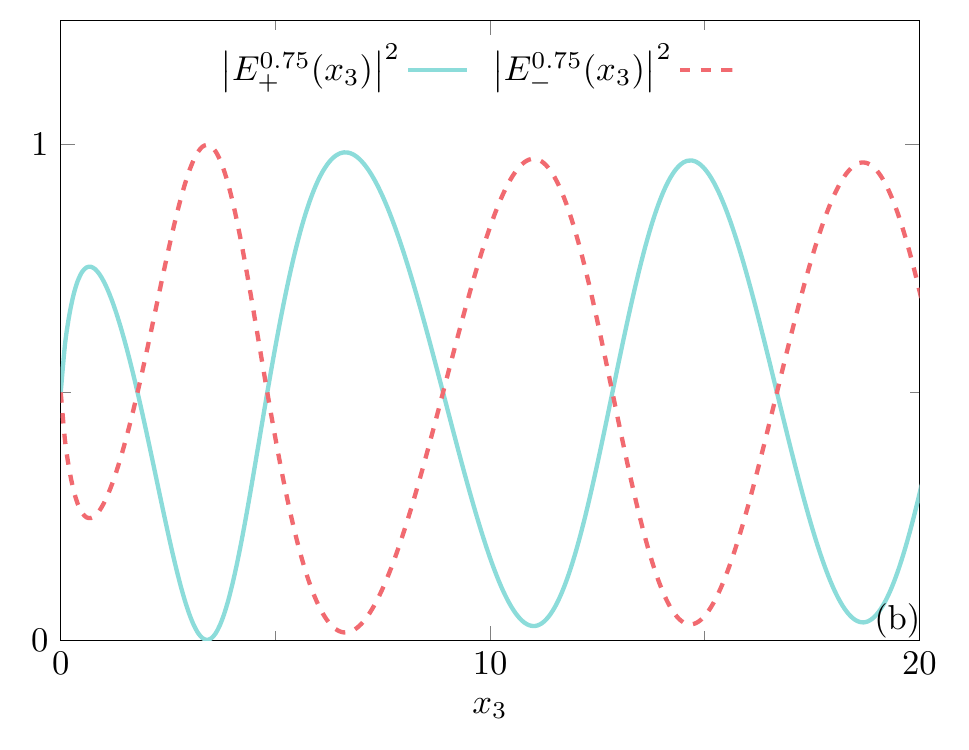}
		\\
		\includegraphics[width=0.45\linewidth]{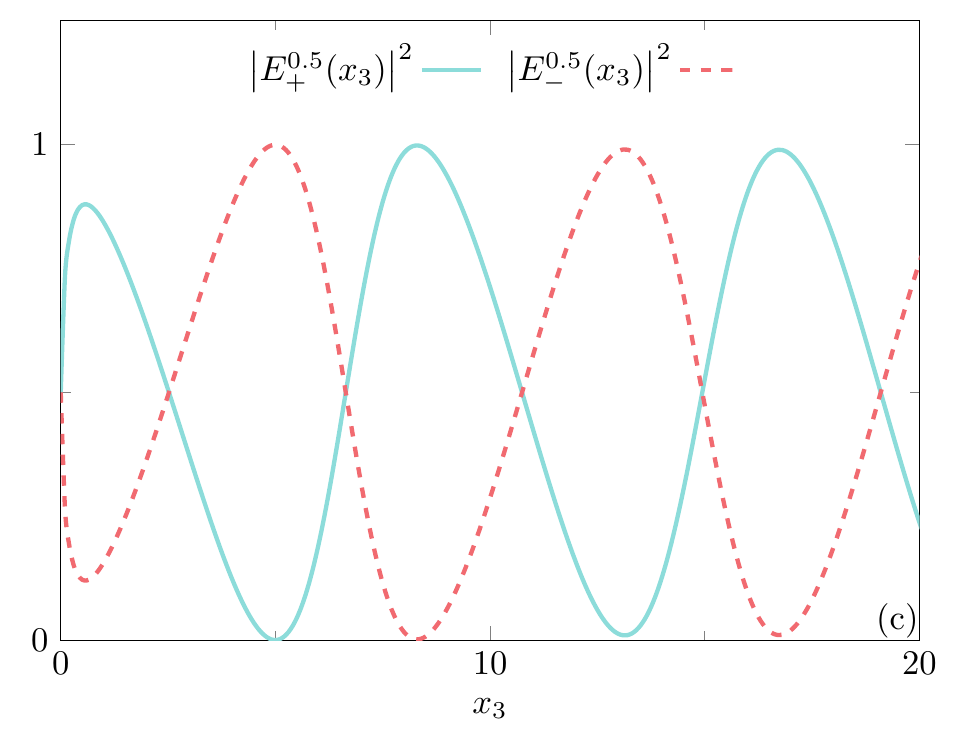}
		\includegraphics[width=0.45\linewidth]{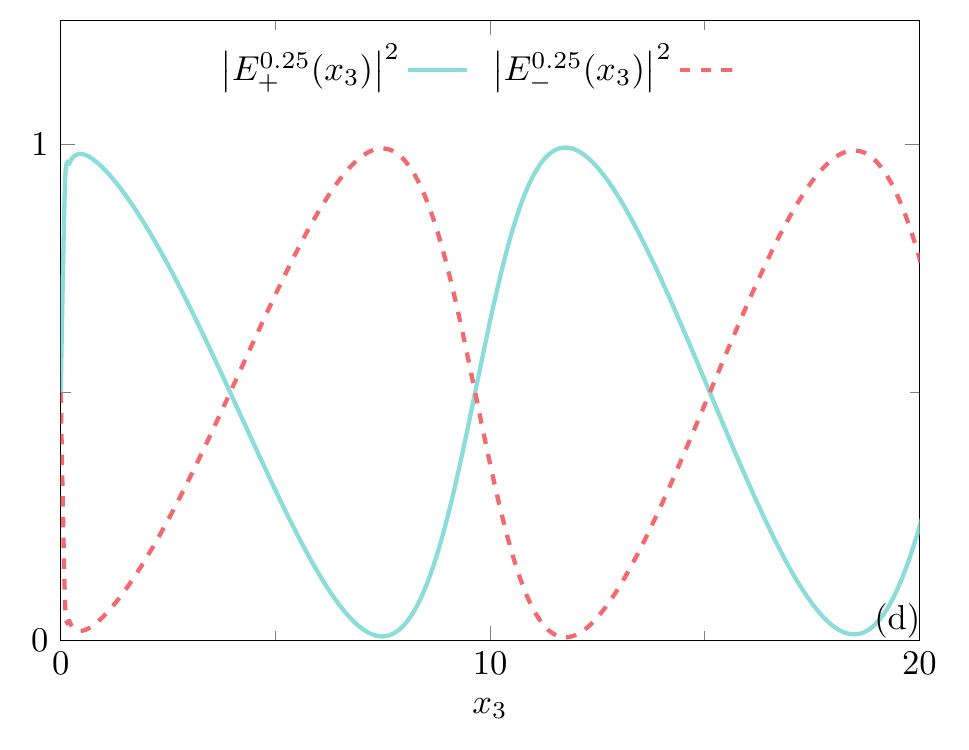}
	\end{center}
	\caption{ \label{fig:EX3} Starting from equal intensity of the fields into the wave guides at $x_{3}=0$, we plot the normalized intensities $\mathrm{I}_{\pm}^{\alpha}(x_3)\sim|E_{\pm}^{\alpha}(x_3)|^{2}$ for the cases: (a) $\alpha=1.0$ (b) $\alpha=0.75$ (c) $\alpha=0.5$ (d) $\alpha=0.25$.}
\end{figure*}

\subsection{Two-coupled wave-guides}
Not only in quantum dynamics, as previously described, the $\mathcal{PT}$-symmetry takes place. In classical optics physics, there is an interesting non-Hermitian $\mathcal{PT}$-symmetric model describing the beam propagation in $\mathcal{PT}$-symmetric complex potential involving coupled waveguides, and which is investigated in Ref. \cite{ruter:10}. In this model, the coupled-mode approach under $\mathcal{PT}$-symmetry implies that the optical-field dynamics, in the two coupled waveguides, is described by the space-dependent Schrödinger-like equations \cite{kahn:17} in the form
\begin{eqnarray*}
    i\frac{d}{dx_3}
    \left[
    \begin{matrix}
    E_{+}(x_3) \\
    E_{-}(x_3)
    \end{matrix}
    \right]
    =
    \left[
    \begin{matrix}
    -i\varepsilon & \varsigma\\
    \varsigma & i\varepsilon
    \end{matrix}
    \right]
    \left[
    \begin{matrix}
    E_{+}(x_3) \\
    E_{-}(x_3)
    \end{matrix}
    \right],
\end{eqnarray*}
where $E_{\pm}(x_3)$ are the field amplitudes in the first and the
second waveguides, and $x_3$ represents the one-dimensional position of
the signals in the waveguides.
The parameters $\varepsilon$ and $\varsigma$ correspond to the effective gain coefficient and the coupling constant. The non-Hermitian ``Hamiltonian'' describing the coupling between the waveguides reads as $\hat{\mathcal{H}}_{0}=\varsigma\hat{\sigma}_{1}-i\varepsilon\hat{\sigma}_{3}$. At this point, the main goal concerns the extension of this model to the fractional scenario paving the way to explore the effects of the fractional derivative on the optical-beam dynamics. This can be done by performing the \textit{ad hoc} extension using the transformations below:
\[
i\frac{d}{dx_3}\rightarrow i^{\alpha}\;^C _0\mathcal{D}_{x_3}^{\alpha},
\quad
E_{\pm}(x_3)\rightarrow E_{\pm}^{\alpha}(x_3),
\quad
\hat{\mathcal{H}}_{0}\rightarrow\hat{\mathcal{H}}_{0}^{\alpha},
\]
from which the ``Hamiltonian'' $\hat{\mathcal{H}}_{0}^{\alpha}$ assumes the form
\begin{eqnarray}
    \hat{\mathcal{H}}_{0}^{\alpha}= \varsigma_{\alpha}\hat{\sigma}_{1}-i \varepsilon_{\alpha}\hat{\sigma}_{3}.
\end{eqnarray}
For simplicity, we consider $\hbar_{\alpha}=1$ and treat all the
parameters as dimensionless.
For this ``Hamiltonian'', we find that the coefficients in
Eq. \eqref{H0} read as $\omega_{1}^{\alpha}=\varsigma_{\alpha}$,
$\omega_{2}^{\alpha}=0$ and $\omega_{3}^{\alpha} =
-i\varepsilon_{\alpha}$, implying
$\Delta_{\alpha} = \sqrt{\varsigma_{\alpha}^{2}-\varepsilon_{\alpha}^{2}}$.
We set the effective gain coefficient and the coupling constant as
$\varepsilon_{\alpha} = 1/2$ and $\varsigma_{\alpha} = 1$, respectively,
with the parameter $\Delta_{\alpha}=\sqrt{3}/2$.
In this case, the dynamics remains into the region of unbroken
$\mathcal{PT}$-symmetry, in which the ``Hamiltonian'' has real
eigenvalues. The behavior of the position-dependent parameters of the
Dyson map (obtained by the change $t\to x_3$ in all our calculations) is
illustrated in Fig. \ref{fig:EX3DysonMap}, assuming the same initial
conditions as done in the previous cases.

We evaluate the intensities of the fields $\mathrm{I}_{\pm}^{\alpha}(x_3)\sim|E_{\pm}^{\alpha}(x_3)|^{2}$ under the unitary time-evolution. Starting from the superposition state $\vert\psi^{\alpha}(0)\rangle = (1,1)^{\mathrm{T}}/\sqrt{2}$, the normalized intensities involve in time in accordance with
\begin{subequations}
\begin{align}
   |E_{\pm}^{\alpha}(x_3)|^{2} &= \frac{1}{2}\left\vert\varpi^{\alpha}(x_3) \pm \tau^{\alpha}(x_3)\right\vert^2.
\end{align}
\end{subequations}
In Fig. \ref{fig:EX3} we plot these intensities for different values of
the fractional parameter $\alpha$.
Notice that at $x_3=0$, the intensities in both waveguides assume the
same value, and along the $x_{3}$-axis,  the intensities of the fields
in both waveguides interchange between them.
It occurs due to the coupling between the waveguides and allows energy
exchange between them.
For $\alpha=1.0$, Fig. \ref{fig:EX3}(a) , there is a small and fast
energy exchange between the waveguides, while for the fractional
parameter $\alpha\neq1$ there is an increase of the field intensities
distributed along the waveguides. 
Moreover, we can see that as the fractionality parameter assumes lower
values, an increasing in the asymmetry of the shape of the energy
distribution appears - small fractionality improves the signal intensity
inside the waveguides indicating an effect of field amplification:
see the lines correspondent to $\alpha=0.75$, Fig.  \ref{fig:EX3}(b),
$\alpha=0.5$, Fig.  \ref{fig:EX3}(c), and $\alpha=0.25$,
Fig. \ref{fig:EX3}(d).


\section{Conclusions}
\label{sec:Conc}
In this work, we discussed different applications in the dynamics of
two-level quantum systems in the fractional-time scenario considering
the Caputo derivative in the context of the formalism developed for the
time-dependent non-Hermitian Hamiltonian operators as proposed in
Ref. \cite{fring:16a,fring:17}.
Both approaches of the fractional-time derivative and the non-Hermitian
Hamiltonian fail to conserve the probability (in relation to the trivial
metric), which is inappropriate from the point of view of the principles
of standard quantum mechanics.
However, our proposed approach allows us to circumvent the non-unitarity
of the time evolution arising from the FTSE, in the well-known formalism
of time-dependent metric as done in non-Hermitian quantum mechanics
\cite{fring:16a,fring:17}.
In applying this formalism, a dynamical Hilbert space with a
time-dependent metric is consistently established, for which the dynamic
proceeds through a unitary evolution.
For generality, we consider the fractional-time dynamics generated by a
non-Hermitian Hamiltonian $\hat{\mathcal{H}}_{0}^{\alpha}$, and some
applications were made in the sense of clarifying the steps of the
approach. 
We observe that the time-dependent Dyson map supplies both the
non-hermiticity coming from $\hat{\mathcal{H}}_{0}^{\alpha}$ as well as
from the fractional-time derivative.
We stress that the parameters are obtained exactly.

Finally, the outlined formalism could be applied to implement a
time-dependent Dyson map \eqref{DysonGD} embedding $SU(1,1)$ generators
in the two-dimensional irreducible representation
$\hat{k}_{0}=\hat{\sigma}_{3}/2$, $\hat{k}_{+}=\hat{\sigma}_{+}$ and
$\hat{k}_{-}=-\hat{\sigma}_{-}$.
Since we have treated a class of discrete finite-dimensional systems, or
more precisely, two-dimensional systems, a natural question arises
concerning analytical solutions for the infinite-dimensional systems,
which we hope to answer in future works.
Our results pave the way for new fractional-time and non-Hermitian
quantum mechanics features.

\section*{Acknowledgements}
D.C. thanks M. H. Y. Moussa for fruitful discussions and the kind
hospitality during the visit to IFSC/USP.
This work was partially supported by the Brazilian agencies Conselho
Nacional de Desenvolvimento Científico e Tecnológico (CNPq) and
Instituto Nacional de Ciência e Tecnologia de Informação Quântica
(INCT-IQ).
It was also financed by the Co\-or\-dena\c{c}\~{a}o de
Aperfei\c{c}oamento de Pessoal de N\'{i}vel Superior (CAPES, Finance
Code 001). 
FMA acknowledges CNPq Grant 314594/2020-5.

\bibliographystyle{apsrev4-2}
\bibliography{references.bib}

\begin{thebibliography}{42}%
\makeatletter
\providecommand \@ifxundefined [1]{%
 \@ifx{#1\undefined}
}%
\providecommand \@ifnum [1]{%
 \ifnum #1\expandafter \@firstoftwo
 \else \expandafter \@secondoftwo
 \fi
}%
\providecommand \@ifx [1]{%
 \ifx #1\expandafter \@firstoftwo
 \else \expandafter \@secondoftwo
 \fi
}%
\providecommand \natexlab [1]{#1}%
\providecommand \enquote  [1]{``#1''}%
\providecommand \bibnamefont  [1]{#1}%
\providecommand \bibfnamefont [1]{#1}%
\providecommand \citenamefont [1]{#1}%
\providecommand \href@noop [0]{\@secondoftwo}%
\providecommand \href [0]{\begingroup \@sanitize@url \@href}%
\providecommand \@href[1]{\@@startlink{#1}\@@href}%
\providecommand \@@href[1]{\endgroup#1\@@endlink}%
\providecommand \@sanitize@url [0]{\catcode `\\12\catcode `\$12\catcode
  `\&12\catcode `\#12\catcode `\^12\catcode `\_12\catcode `\%12\relax}%
\providecommand \@@startlink[1]{}%
\providecommand \@@endlink[0]{}%
\providecommand \url  [0]{\begingroup\@sanitize@url \@url }%
\providecommand \@url [1]{\endgroup\@href {#1}{\urlprefix }}%
\providecommand \urlprefix  [0]{URL }%
\providecommand \Eprint [0]{\href }%
\providecommand \doibase [0]{https://doi.org/}%
\providecommand \selectlanguage [0]{\@gobble}%
\providecommand \bibinfo  [0]{\@secondoftwo}%
\providecommand \bibfield  [0]{\@secondoftwo}%
\providecommand \translation [1]{[#1]}%
\providecommand \BibitemOpen [0]{}%
\providecommand \bibitemStop [0]{}%
\providecommand \bibitemNoStop [0]{.\EOS\space}%
\providecommand \EOS [0]{\spacefactor3000\relax}%
\providecommand \BibitemShut  [1]{\csname bibitem#1\endcsname}%
\let\auto@bib@innerbib\@empty
\bibitem [{\citenamefont {Longhi}(2015)}]{longhi2015fractional}%
  \BibitemOpen
  \bibfield  {author} {\bibinfo {author} {\bibfnamefont {S.}~\bibnamefont
  {Longhi}},\ }\href {https://doi.org/https://doi.org/10.1364/OL.40.001117}
  {\bibfield  {journal} {\bibinfo  {journal} {Opt. Lett}\ }\textbf {\bibinfo
  {volume} {40}},\ \bibinfo {pages} {1117} (\bibinfo {year}
  {2015})}\BibitemShut {NoStop}%
\bibitem [{\citenamefont {Huang}\ and\ \citenamefont
  {Dong}(2017)}]{huang2017beam}%
  \BibitemOpen
  \bibfield  {author} {\bibinfo {author} {\bibfnamefont {C.}~\bibnamefont
  {Huang}}\ and\ \bibinfo {author} {\bibfnamefont {L.}~\bibnamefont {Dong}},\
  }\href {https://doi.org/https://doi.org/10.1038/s41598-017-05926-5}
  {\bibfield  {journal} {\bibinfo  {journal} {Sci. Rep.}\ }\textbf {\bibinfo
  {volume} {7}},\ \bibinfo {pages} {1} (\bibinfo {year} {2017})}\BibitemShut
  {NoStop}%
\bibitem [{\citenamefont {Zhang}\ \emph {et~al.}(2016)\citenamefont {Zhang},
  \citenamefont {Zhong}, \citenamefont {Beli{\'c}}, \citenamefont {Zhu},
  \citenamefont {Zhong}, \citenamefont {Zhang}, \citenamefont
  {Christodoulides},\ and\ \citenamefont {Xiao}}]{zhang2016pt}%
  \BibitemOpen
  \bibfield  {author} {\bibinfo {author} {\bibfnamefont {Y.}~\bibnamefont
  {Zhang}}, \bibinfo {author} {\bibfnamefont {H.}~\bibnamefont {Zhong}},
  \bibinfo {author} {\bibfnamefont {M.~R.}\ \bibnamefont {Beli{\'c}}}, \bibinfo
  {author} {\bibfnamefont {Y.}~\bibnamefont {Zhu}}, \bibinfo {author}
  {\bibfnamefont {W.}~\bibnamefont {Zhong}}, \bibinfo {author} {\bibfnamefont
  {Y.}~\bibnamefont {Zhang}}, \bibinfo {author} {\bibfnamefont {D.~N.}\
  \bibnamefont {Christodoulides}},\ and\ \bibinfo {author} {\bibfnamefont
  {M.}~\bibnamefont {Xiao}},\ }\href
  {https://doi.org/https://doi.org/10.1002/lpor.201600037} {\bibfield
  {journal} {\bibinfo  {journal} {Laser Photonics Rev.}\ }\textbf {\bibinfo
  {volume} {10}},\ \bibinfo {pages} {526} (\bibinfo {year} {2016})}\BibitemShut
  {NoStop}%
\bibitem [{\citenamefont {Heydari}\ and\ \citenamefont
  {Atangana}(2019)}]{heydari2019cardinal}%
  \BibitemOpen
  \bibfield  {author} {\bibinfo {author} {\bibfnamefont {M.}~\bibnamefont
  {Heydari}}\ and\ \bibinfo {author} {\bibfnamefont {A.}~\bibnamefont
  {Atangana}},\ }\href
  {https://doi.org/https://doi.org/10.1016/j.chaos.2019.08.009} {\bibfield
  {journal} {\bibinfo  {journal} {Chaos Solit. Fractals}\ }\textbf {\bibinfo
  {volume} {128}},\ \bibinfo {pages} {339} (\bibinfo {year}
  {2019})}\BibitemShut {NoStop}%
\bibitem [{\citenamefont {Wu}\ \emph {et~al.}(2010)\citenamefont {Wu},
  \citenamefont {Huang}, \citenamefont {Cheng},\ and\ \citenamefont
  {Hsieh}}]{wu2010}%
  \BibitemOpen
  \bibfield  {author} {\bibinfo {author} {\bibfnamefont {J.-N.}\ \bibnamefont
  {Wu}}, \bibinfo {author} {\bibfnamefont {C.-H.}\ \bibnamefont {Huang}},
  \bibinfo {author} {\bibfnamefont {S.-C.}\ \bibnamefont {Cheng}},\ and\
  \bibinfo {author} {\bibfnamefont {W.-F.}\ \bibnamefont {Hsieh}},\ }\href
  {https://doi.org/10.1103/PhysRevA.81.023827} {\bibfield  {journal} {\bibinfo
  {journal} {Phys. Rev. A}\ }\textbf {\bibinfo {volume} {81}},\ \bibinfo
  {pages} {023827} (\bibinfo {year} {2010})}\BibitemShut {NoStop}%
\bibitem [{\citenamefont {Fujita}\ \emph {et~al.}(2005)\citenamefont {Fujita},
  \citenamefont {Takahashi}, \citenamefont {Tanaka}, \citenamefont {Asano},\
  and\ \citenamefont {Noda}}]{fujita2005}%
  \BibitemOpen
  \bibfield  {author} {\bibinfo {author} {\bibfnamefont {M.}~\bibnamefont
  {Fujita}}, \bibinfo {author} {\bibfnamefont {S.}~\bibnamefont {Takahashi}},
  \bibinfo {author} {\bibfnamefont {Y.}~\bibnamefont {Tanaka}}, \bibinfo
  {author} {\bibfnamefont {T.}~\bibnamefont {Asano}},\ and\ \bibinfo {author}
  {\bibfnamefont {S.}~\bibnamefont {Noda}},\ }\href
  {https://doi.org/10.1126/science.1110417} {\bibfield  {journal} {\bibinfo
  {journal} {Science}\ }\textbf {\bibinfo {volume} {308}},\ \bibinfo {pages}
  {1296} (\bibinfo {year} {2005})}\BibitemShut {NoStop}%
\bibitem [{\citenamefont {Podlubny}(1998)}]{podlubny1998fractional}%
  \BibitemOpen
  \bibfield  {author} {\bibinfo {author} {\bibfnamefont {I.}~\bibnamefont
  {Podlubny}},\ }\href@noop {} {\emph {\bibinfo {title} {Fractional
  differential equations}}}\ (\bibinfo  {publisher} {Academic Press},\ \bibinfo
  {year} {1998})\BibitemShut {NoStop}%
\bibitem [{\citenamefont {Metzler}\ and\ \citenamefont
  {Klafter}(2000)}]{metzler2000random}%
  \BibitemOpen
  \bibfield  {author} {\bibinfo {author} {\bibfnamefont {R.}~\bibnamefont
  {Metzler}}\ and\ \bibinfo {author} {\bibfnamefont {J.}~\bibnamefont
  {Klafter}},\ }\href
  {https://doi.org/https://doi.org/10.1016/S0370-1573(00)00070-3} {\bibfield
  {journal} {\bibinfo  {journal} {Phys. Rep.}\ }\textbf {\bibinfo {volume}
  {339}},\ \bibinfo {pages} {1} (\bibinfo {year} {2000})}\BibitemShut {NoStop}%
\bibitem [{\citenamefont {Laskin}(2000{\natexlab{a}})}]{laskin:00a}%
  \BibitemOpen
  \bibfield  {author} {\bibinfo {author} {\bibfnamefont {N.}~\bibnamefont
  {Laskin}},\ }\href
  {https://doi.org/https://doi.org/10.1016/S0375-9601(00)00201-2} {\bibfield
  {journal} {\bibinfo  {journal} {Phys. Lett. A}\ }\textbf {\bibinfo {volume}
  {268}},\ \bibinfo {pages} {298} (\bibinfo {year}
  {2000}{\natexlab{a}})}\BibitemShut {NoStop}%
\bibitem [{\citenamefont {Laskin}(2000{\natexlab{b}})}]{laskin:00b}%
  \BibitemOpen
  \bibfield  {author} {\bibinfo {author} {\bibfnamefont {N.}~\bibnamefont
  {Laskin}},\ }\href {https://doi.org/https://doi.org/10.1103/PhysRevE.62.3135}
  {\bibfield  {journal} {\bibinfo  {journal} {Phys. Rev. E}\ }\textbf {\bibinfo
  {volume} {62}},\ \bibinfo {pages} {3135} (\bibinfo {year}
  {2000}{\natexlab{b}})}\BibitemShut {NoStop}%
\bibitem [{\citenamefont {Wei}(2016)}]{wei:16}%
  \BibitemOpen
  \bibfield  {author} {\bibinfo {author} {\bibfnamefont {Y.}~\bibnamefont
  {Wei}},\ }\href {https://doi.org/https://doi.org/10.1103/PhysRevE.93.066103}
  {\bibfield  {journal} {\bibinfo  {journal} {Phys. Rev. E}\ }\textbf {\bibinfo
  {volume} {93}},\ \bibinfo {pages} {066103} (\bibinfo {year}
  {2016})}\BibitemShut {NoStop}%
\bibitem [{\citenamefont {Laskin}(2016)}]{laskin:16}%
  \BibitemOpen
  \bibfield  {author} {\bibinfo {author} {\bibfnamefont {N.}~\bibnamefont
  {Laskin}},\ }\href
  {https://doi.org/https://doi.org/10.1103/PhysRevE.93.066104} {\bibfield
  {journal} {\bibinfo  {journal} {Phys. Rev. E}\ }\textbf {\bibinfo {volume}
  {93}},\ \bibinfo {pages} {066104} (\bibinfo {year} {2016})}\BibitemShut
  {NoStop}%
\bibitem [{\citenamefont {Naber}(2004)}]{naber:04}%
  \BibitemOpen
  \bibfield  {author} {\bibinfo {author} {\bibfnamefont {M.}~\bibnamefont
  {Naber}},\ }\href {https://doi.org/https://doi.org/10.1063/1.1769611}
  {\bibfield  {journal} {\bibinfo  {journal} {J. Math. Phys.}\ }\textbf
  {\bibinfo {volume} {45}},\ \bibinfo {pages} {3339} (\bibinfo {year}
  {2004})}\BibitemShut {NoStop}%
\bibitem [{\citenamefont {Iomin}(2009)}]{iomin2009fractional}%
  \BibitemOpen
  \bibfield  {author} {\bibinfo {author} {\bibfnamefont {A.}~\bibnamefont
  {Iomin}},\ }\href
  {https://doi.org/https://doi.org/10.1103/PhysRevE.80.022103} {\bibfield
  {journal} {\bibinfo  {journal} {Phys. Rev. E}\ }\textbf {\bibinfo {volume}
  {80}},\ \bibinfo {pages} {022103} (\bibinfo {year} {2009})}\BibitemShut
  {NoStop}%
\bibitem [{\citenamefont {Nasrolahpour}(2011)}]{nasrolahpour2011electron}%
  \BibitemOpen
  \bibfield  {author} {\bibinfo {author} {\bibfnamefont {H.}~\bibnamefont
  {Nasrolahpour}},\ }\href@noop {} {\bibfield  {journal} {\bibinfo  {journal}
  {Prespacetime Journal}\ }\textbf {\bibinfo {volume} {2}},\ \bibinfo {pages}
  {2053} (\bibinfo {year} {2011})}\BibitemShut {NoStop}%
\bibitem [{\citenamefont {Lenzi}\ \emph {et~al.}(2013)\citenamefont {Lenzi},
  \citenamefont {Ribeiro}, \citenamefont {dos Santos}, \citenamefont
  {Rossato},\ and\ \citenamefont {Mendes}}]{lenzi2013time}%
  \BibitemOpen
  \bibfield  {author} {\bibinfo {author} {\bibfnamefont {E.~K.}\ \bibnamefont
  {Lenzi}}, \bibinfo {author} {\bibfnamefont {H.~V.}\ \bibnamefont {Ribeiro}},
  \bibinfo {author} {\bibfnamefont {M.~A.~F.}\ \bibnamefont {dos Santos}},
  \bibinfo {author} {\bibfnamefont {R.}~\bibnamefont {Rossato}},\ and\ \bibinfo
  {author} {\bibfnamefont {R.~S.}\ \bibnamefont {Mendes}},\ }\href
  {https://doi.org/https://doi.org/10.1063/1.4819253} {\bibfield  {journal}
  {\bibinfo  {journal} {J. Math. Phys.}\ }\textbf {\bibinfo {volume} {54}},\
  \bibinfo {pages} {082107} (\bibinfo {year} {2013})}\BibitemShut {NoStop}%
\bibitem [{\citenamefont {Iomin}(2020)}]{iomin2020}%
  \BibitemOpen
  \bibfield  {author} {\bibinfo {author} {\bibfnamefont {A.}~\bibnamefont
  {Iomin}},\ }\href
  {https://doi.org/https://doi.org/10.1016/j.chaos.2020.110305} {\bibfield
  {journal} {\bibinfo  {journal} {Chaos Solit. Fractals}\ }\textbf {\bibinfo
  {volume} {139}},\ \bibinfo {pages} {110305} (\bibinfo {year}
  {2020})}\BibitemShut {NoStop}%
\bibitem [{\citenamefont {Laskin}(2017)}]{laskin2017time}%
  \BibitemOpen
  \bibfield  {author} {\bibinfo {author} {\bibfnamefont {N.}~\bibnamefont
  {Laskin}},\ }\href
  {https://doi.org/https://doi.org/10.1016/j.chaos.2017.04.010} {\bibfield
  {journal} {\bibinfo  {journal} {Chaos Solit. Fractals}\ }\textbf {\bibinfo
  {volume} {102}},\ \bibinfo {pages} {16} (\bibinfo {year} {2017})}\BibitemShut
  {NoStop}%
\bibitem [{\citenamefont {Iomin}(2019{\natexlab{a}})}]{iomin2019app}%
  \BibitemOpen
  \bibfield  {author} {\bibinfo {author} {\bibfnamefont {A.}~\bibnamefont
  {Iomin}},\ }\bibinfo {title} {Fractional time quantum mechanics},\ in\ \href
  {https://doi.org/doi:10.1515/9783110571721-013} {\emph {\bibinfo {booktitle}
  {Volume 5 Applications in Physics, Part B}}},\ \bibinfo {editor} {edited by\
  \bibinfo {editor} {\bibfnamefont {V.~E.}\ \bibnamefont {Tarasov}}}\ (\bibinfo
   {publisher} {De Gruyter},\ \bibinfo {address} {Berlin, Boston},\ \bibinfo
  {year} {2019})\ pp.\ \bibinfo {pages} {299--316}\BibitemShut {NoStop}%
\bibitem [{\citenamefont {Zhang}\ \emph {et~al.}(2021)\citenamefont {Zhang},
  \citenamefont {Yang}, \citenamefont {Wei},\ and\ \citenamefont
  {Luo}}]{zhang2021quantization}%
  \BibitemOpen
  \bibfield  {author} {\bibinfo {author} {\bibfnamefont {X.}~\bibnamefont
  {Zhang}}, \bibinfo {author} {\bibfnamefont {B.}~\bibnamefont {Yang}},
  \bibinfo {author} {\bibfnamefont {C.}~\bibnamefont {Wei}},\ and\ \bibinfo
  {author} {\bibfnamefont {M.}~\bibnamefont {Luo}},\ }\href
  {https://doi.org/https://doi.org/10.1016/j.cnsns.2020.105531} {\bibfield
  {journal} {\bibinfo  {journal} {Commun. Nonlinear Sci. Numer. Simul.}\
  }\textbf {\bibinfo {volume} {93}},\ \bibinfo {pages} {105531} (\bibinfo
  {year} {2021})}\BibitemShut {NoStop}%
\bibitem [{\citenamefont {Iomin}(2019{\natexlab{b}})}]{iomin2019fractional}%
  \BibitemOpen
  \bibfield  {author} {\bibinfo {author} {\bibfnamefont {A.}~\bibnamefont
  {Iomin}},\ }\href
  {https://doi.org/https://doi.org/10.1016/j.csfx.2018.100001} {\bibfield
  {journal} {\bibinfo  {journal} {Chaos Solit. Fractals: X}\ }\textbf {\bibinfo
  {volume} {1}},\ \bibinfo {pages} {100001} (\bibinfo {year}
  {2019}{\natexlab{b}})}\BibitemShut {NoStop}%
\bibitem [{\citenamefont {Bender}\ and\ \citenamefont
  {Boettcher}(1998)}]{bender:98}%
  \BibitemOpen
  \bibfield  {author} {\bibinfo {author} {\bibfnamefont {C.~M.}\ \bibnamefont
  {Bender}}\ and\ \bibinfo {author} {\bibfnamefont {S.}~\bibnamefont
  {Boettcher}},\ }\href
  {https://doi.org/https://doi.org/10.1103/PhysRevLett.80.5243} {\bibfield
  {journal} {\bibinfo  {journal} {Phys. Rev. Lett.}\ }\textbf {\bibinfo
  {volume} {80}},\ \bibinfo {pages} {5243} (\bibinfo {year}
  {1998})}\BibitemShut {NoStop}%
\bibitem [{\citenamefont {Bender}(2007)}]{bender:07}%
  \BibitemOpen
  \bibfield  {author} {\bibinfo {author} {\bibfnamefont {C.~M.}\ \bibnamefont
  {Bender}},\ }\href
  {https://doi.org/https://doi.org/10.1088/0034-4885/70/6/R03} {\bibfield
  {journal} {\bibinfo  {journal} {Rep. Prog. Phys.}\ }\textbf {\bibinfo
  {volume} {70}},\ \bibinfo {pages} {947} (\bibinfo {year} {2007})}\BibitemShut
  {NoStop}%
\bibitem [{\citenamefont {Dorey}\ \emph {et~al.}(2001)\citenamefont {Dorey},
  \citenamefont {Dunning},\ and\ \citenamefont {Tateo}}]{dorey:01}%
  \BibitemOpen
  \bibfield  {author} {\bibinfo {author} {\bibfnamefont {P.}~\bibnamefont
  {Dorey}}, \bibinfo {author} {\bibfnamefont {C.}~\bibnamefont {Dunning}},\
  and\ \bibinfo {author} {\bibfnamefont {R.}~\bibnamefont {Tateo}},\ }\href
  {https://doi.org/https://doi.org/10.1088/0305-4470/34/28/305} {\bibfield
  {journal} {\bibinfo  {journal} {J. Phys. A}\ }\textbf {\bibinfo {volume}
  {34}},\ \bibinfo {pages} {5679} (\bibinfo {year} {2001})}\BibitemShut
  {NoStop}%
\bibitem [{\citenamefont
  {Mostafazadeh}(2002{\natexlab{a}})}]{mostafazadeh:02a}%
  \BibitemOpen
  \bibfield  {author} {\bibinfo {author} {\bibfnamefont {A.}~\bibnamefont
  {Mostafazadeh}},\ }\href {https://doi.org/https://doi.org/10.1063/1.1418246}
  {\bibfield  {journal} {\bibinfo  {journal} {J. Math. Phys.}\ }\textbf
  {\bibinfo {volume} {43}},\ \bibinfo {pages} {205} (\bibinfo {year}
  {2002}{\natexlab{a}})}\BibitemShut {NoStop}%
\bibitem [{\citenamefont
  {Mostafazadeh}(2002{\natexlab{b}})}]{mostafazadeh:02b}%
  \BibitemOpen
  \bibfield  {author} {\bibinfo {author} {\bibfnamefont {A.}~\bibnamefont
  {Mostafazadeh}},\ }\href {https://doi.org/https://doi.org/10.1063/1.1461427}
  {\bibfield  {journal} {\bibinfo  {journal} {J. Math. Phys.}\ }\textbf
  {\bibinfo {volume} {43}},\ \bibinfo {pages} {2814} (\bibinfo {year}
  {2002}{\natexlab{b}})}\BibitemShut {NoStop}%
\bibitem [{\citenamefont
  {Mostafazadeh}(2002{\natexlab{c}})}]{mostafazadeh:02c}%
  \BibitemOpen
  \bibfield  {author} {\bibinfo {author} {\bibfnamefont {A.}~\bibnamefont
  {Mostafazadeh}},\ }\href {https://doi.org/https://doi.org/10.1063/1.1489072}
  {\bibfield  {journal} {\bibinfo  {journal} {J. Math. Phys.}\ }\textbf
  {\bibinfo {volume} {43}},\ \bibinfo {pages} {3944} (\bibinfo {year}
  {2002}{\natexlab{c}})}\BibitemShut {NoStop}%
\bibitem [{\citenamefont {Scholtz}\ \emph {et~al.}(1992)\citenamefont
  {Scholtz}, \citenamefont {Geyer},\ and\ \citenamefont {Hahne}}]{scholtz:92}%
  \BibitemOpen
  \bibfield  {author} {\bibinfo {author} {\bibfnamefont {F.}~\bibnamefont
  {Scholtz}}, \bibinfo {author} {\bibfnamefont {H.}~\bibnamefont {Geyer}},\
  and\ \bibinfo {author} {\bibfnamefont {F.}~\bibnamefont {Hahne}},\ }\href
  {https://doi.org/https://doi.org/10.1016/0003-4916(92)90284-S} {\bibfield
  {journal} {\bibinfo  {journal} {Ann. Phys.}\ }\textbf {\bibinfo {volume}
  {213}},\ \bibinfo {pages} {74} (\bibinfo {year} {1992})}\BibitemShut
  {NoStop}%
\bibitem [{\citenamefont {Znojil}(2008)}]{znojil:08}%
  \BibitemOpen
  \bibfield  {author} {\bibinfo {author} {\bibfnamefont {M.}~\bibnamefont
  {Znojil}},\ }\href
  {https://doi.org/https://doi.org/10.1103/PhysRevD.78.085003} {\bibfield
  {journal} {\bibinfo  {journal} {Phys. Rev. D}\ }\textbf {\bibinfo {volume}
  {78}},\ \bibinfo {pages} {085003} (\bibinfo {year} {2008})}\BibitemShut
  {NoStop}%
\bibitem [{\citenamefont {Gong}\ and\ \citenamefont {Wang}(2013)}]{gong:13}%
  \BibitemOpen
  \bibfield  {author} {\bibinfo {author} {\bibfnamefont {J.}~\bibnamefont
  {Gong}}\ and\ \bibinfo {author} {\bibfnamefont {Q.}~\bibnamefont {Wang}},\
  }\href {https://doi.org/https://doi.org/10.1088/1751-8113/46/48/485302}
  {\bibfield  {journal} {\bibinfo  {journal} {J. Phys. A}\ }\textbf {\bibinfo
  {volume} {46}},\ \bibinfo {pages} {485302} (\bibinfo {year}
  {2013})}\BibitemShut {NoStop}%
\bibitem [{\citenamefont {Fring}\ and\ \citenamefont
  {Moussa}(2016)}]{fring:16a}%
  \BibitemOpen
  \bibfield  {author} {\bibinfo {author} {\bibfnamefont {A.}~\bibnamefont
  {Fring}}\ and\ \bibinfo {author} {\bibfnamefont {M.~H.~Y.}\ \bibnamefont
  {Moussa}},\ }\href
  {https://doi.org/https://doi.org/10.1103/PhysRevA.93.042114} {\bibfield
  {journal} {\bibinfo  {journal} {Phys. Rev. A}\ }\textbf {\bibinfo {volume}
  {93}},\ \bibinfo {pages} {042114} (\bibinfo {year} {2016})}\BibitemShut
  {NoStop}%
\bibitem [{\citenamefont {Luiz}\ \emph {et~al.}(2020)\citenamefont {Luiz},
  \citenamefont {de~Ponte},\ and\ \citenamefont {Moussa}}]{luiz:20}%
  \BibitemOpen
  \bibfield  {author} {\bibinfo {author} {\bibfnamefont {F.~S.}\ \bibnamefont
  {Luiz}}, \bibinfo {author} {\bibfnamefont {M.~A.}\ \bibnamefont {de~Ponte}},\
  and\ \bibinfo {author} {\bibfnamefont {M.~H.~Y.}\ \bibnamefont {Moussa}},\
  }\href {https://doi.org/https://doi.org/10.1088/1402-4896/ab80e5} {\bibfield
  {journal} {\bibinfo  {journal} {Phys. Scr.}\ }\textbf {\bibinfo {volume}
  {95}},\ \bibinfo {pages} {065211} (\bibinfo {year} {2020})}\BibitemShut
  {NoStop}%
\bibitem [{\citenamefont {Fring}\ and\ \citenamefont {Frith}(2017)}]{fring:17}%
  \BibitemOpen
  \bibfield  {author} {\bibinfo {author} {\bibfnamefont {A.}~\bibnamefont
  {Fring}}\ and\ \bibinfo {author} {\bibfnamefont {T.}~\bibnamefont {Frith}},\
  }\href {https://doi.org/https://doi.org/10.1016/j.physleta.2017.05.041}
  {\bibfield  {journal} {\bibinfo  {journal} {Phys. Lett. A}\ }\textbf
  {\bibinfo {volume} {381}},\ \bibinfo {pages} {2318} (\bibinfo {year}
  {2017})}\BibitemShut {NoStop}%
\bibitem [{\citenamefont {Maamache}\ \emph {et~al.}(2017)\citenamefont
  {Maamache}, \citenamefont {Djeghiour}, \citenamefont {Mana},\ and\
  \citenamefont {Koussa}}]{maamache:17}%
  \BibitemOpen
  \bibfield  {author} {\bibinfo {author} {\bibfnamefont {M.}~\bibnamefont
  {Maamache}}, \bibinfo {author} {\bibfnamefont {O.~K.}\ \bibnamefont
  {Djeghiour}}, \bibinfo {author} {\bibfnamefont {N.}~\bibnamefont {Mana}},\
  and\ \bibinfo {author} {\bibfnamefont {W.}~\bibnamefont {Koussa}},\ }\href
  {https://doi.org/https://doi.org/10.1140/epjp/i2017-11678-2} {\bibfield
  {journal} {\bibinfo  {journal} {Eur. Phys. J. Plus}\ }\textbf {\bibinfo
  {volume} {132}},\ \bibinfo {pages} {1} (\bibinfo {year} {2017})}\BibitemShut
  {NoStop}%
\bibitem [{\citenamefont {Fring}\ and\ \citenamefont {Frith}(2019)}]{fring:19}%
  \BibitemOpen
  \bibfield  {author} {\bibinfo {author} {\bibfnamefont {A.}~\bibnamefont
  {Fring}}\ and\ \bibinfo {author} {\bibfnamefont {T.}~\bibnamefont {Frith}},\
  }\href {https://doi.org/https://doi.org/10.1103/PhysRevA.100.010102}
  {\bibfield  {journal} {\bibinfo  {journal} {Phys. Rev. A}\ }\textbf {\bibinfo
  {volume} {100}},\ \bibinfo {pages} {010102} (\bibinfo {year}
  {2019})}\BibitemShut {NoStop}%
\bibitem [{\citenamefont {Khantoul}\ \emph {et~al.}(2017)\citenamefont
  {Khantoul}, \citenamefont {Bounames},\ and\ \citenamefont
  {Maamache}}]{khantoul:17}%
  \BibitemOpen
  \bibfield  {author} {\bibinfo {author} {\bibfnamefont {B.}~\bibnamefont
  {Khantoul}}, \bibinfo {author} {\bibfnamefont {A.}~\bibnamefont {Bounames}},\
  and\ \bibinfo {author} {\bibfnamefont {M.}~\bibnamefont {Maamache}},\ }\href
  {https://doi.org/https://doi.org/10.1140/epjp/i2017-11524-7} {\bibfield
  {journal} {\bibinfo  {journal} {Eur. Phys. J. Plus}\ }\textbf {\bibinfo
  {volume} {132}},\ \bibinfo {pages} {1} (\bibinfo {year} {2017})}\BibitemShut
  {NoStop}%
\bibitem [{\citenamefont {Mana}\ \emph {et~al.}(2020)\citenamefont {Mana},
  \citenamefont {Zaidi},\ and\ \citenamefont {Maamache}}]{mana:20}%
  \BibitemOpen
  \bibfield  {author} {\bibinfo {author} {\bibfnamefont {N.}~\bibnamefont
  {Mana}}, \bibinfo {author} {\bibfnamefont {O.}~\bibnamefont {Zaidi}},\ and\
  \bibinfo {author} {\bibfnamefont {M.}~\bibnamefont {Maamache}},\ }\href
  {https://doi.org/https://doi.org/10.1063/5.0013723} {\bibfield  {journal}
  {\bibinfo  {journal} {J. Math. Phys.}\ }\textbf {\bibinfo {volume} {61}},\
  \bibinfo {pages} {102103} (\bibinfo {year} {2020})}\BibitemShut {NoStop}%
\bibitem [{\citenamefont {Koussa}\ \emph {et~al.}(2020)\citenamefont {Koussa},
  \citenamefont {Attia},\ and\ \citenamefont {Maamache}}]{koussa:20}%
  \BibitemOpen
  \bibfield  {author} {\bibinfo {author} {\bibfnamefont {W.}~\bibnamefont
  {Koussa}}, \bibinfo {author} {\bibfnamefont {M.}~\bibnamefont {Attia}},\ and\
  \bibinfo {author} {\bibfnamefont {M.}~\bibnamefont {Maamache}},\ }\href
  {https://doi.org/https://doi.org/10.1063/1.5145269} {\bibfield  {journal}
  {\bibinfo  {journal} {J. Math. Phys.}\ }\textbf {\bibinfo {volume} {61}},\
  \bibinfo {pages} {042101} (\bibinfo {year} {2020})}\BibitemShut {NoStop}%
\bibitem [{\citenamefont {Cius}\ \emph {et~al.}(2022)\citenamefont {Cius},
  \citenamefont {Andrade}, \citenamefont {{de Castro}},\ and\ \citenamefont
  {Moussa}}]{cius:22}%
  \BibitemOpen
  \bibfield  {author} {\bibinfo {author} {\bibfnamefont {D.}~\bibnamefont
  {Cius}}, \bibinfo {author} {\bibfnamefont {F.}~\bibnamefont {Andrade}},
  \bibinfo {author} {\bibfnamefont {A.}~\bibnamefont {{de Castro}}},\ and\
  \bibinfo {author} {\bibfnamefont {M.}~\bibnamefont {Moussa}},\ }\href
  {https://doi.org/https://doi.org/10.1016/j.physa.2022.126945} {\bibfield
  {journal} {\bibinfo  {journal} {Phys. A: Stat. Mech. Appl.}\ }\textbf
  {\bibinfo {volume} {593}},\ \bibinfo {pages} {126945} (\bibinfo {year}
  {2022})}\BibitemShut {NoStop}%
\bibitem [{\citenamefont {von Gehlen}(1991)}]{gehlen:91}%
  \BibitemOpen
  \bibfield  {author} {\bibinfo {author} {\bibfnamefont {G.}~\bibnamefont {von
  Gehlen}},\ }\href
  {https://doi.org/https://doi.org/10.1088/0305-4470/24/22/021} {\bibfield
  {journal} {\bibinfo  {journal} {J. Phys. A}\ }\textbf {\bibinfo {volume}
  {24}},\ \bibinfo {pages} {5371} (\bibinfo {year} {1991})}\BibitemShut
  {NoStop}%
\bibitem [{\citenamefont {R{\"u}ter}\ \emph {et~al.}(2010)\citenamefont
  {R{\"u}ter}, \citenamefont {Makris}, \citenamefont {El-Ganainy},
  \citenamefont {Christodoulides}, \citenamefont {Segev},\ and\ \citenamefont
  {Kip}}]{ruter:10}%
  \BibitemOpen
  \bibfield  {author} {\bibinfo {author} {\bibfnamefont {C.~E.}\ \bibnamefont
  {R{\"u}ter}}, \bibinfo {author} {\bibfnamefont {K.~G.}\ \bibnamefont
  {Makris}}, \bibinfo {author} {\bibfnamefont {R.}~\bibnamefont {El-Ganainy}},
  \bibinfo {author} {\bibfnamefont {D.~N.}\ \bibnamefont {Christodoulides}},
  \bibinfo {author} {\bibfnamefont {M.}~\bibnamefont {Segev}},\ and\ \bibinfo
  {author} {\bibfnamefont {D.}~\bibnamefont {Kip}},\ }\href
  {https://doi.org/https://doi.org/10.1038/nphys1515} {\bibfield  {journal}
  {\bibinfo  {journal} {Nat. Phys.}\ }\textbf {\bibinfo {volume} {6}},\
  \bibinfo {pages} {192} (\bibinfo {year} {2010})}\BibitemShut {NoStop}%
\bibitem [{\citenamefont {Kahn}\ and\ \citenamefont {Marcus}(2017)}]{kahn:17}%
  \BibitemOpen
  \bibfield  {author} {\bibinfo {author} {\bibfnamefont {M.}~\bibnamefont
  {Kahn}}\ and\ \bibinfo {author} {\bibfnamefont {G.}~\bibnamefont {Marcus}},\
  }\href {https://doi.org/https://doi.org/10.1088/1361-6455/aa65a9} {\bibfield
  {journal} {\bibinfo  {journal} {J. Phys. B}\ }\textbf {\bibinfo {volume}
  {50}},\ \bibinfo {pages} {095004} (\bibinfo {year} {2017})}\BibitemShut
  {NoStop}%
\end{thebibliography}%

\end{document}